\documentclass[10pt]{revtex4} 
\usepackage{amsmath}
\usepackage{amssymb}
\usepackage{amsthm}
\usepackage{amsfonts}
\usepackage[latin1]{inputenc}
\usepackage{graphicx}

\newtheorem{lem}{Lemma}

\newtheorem{thm}{Theorem}
\newtheorem*{thm*}{Theorem}

\numberwithin{equation}{section}

\def\pp{{\bf p}}

\def\xx{{\bf x}}
\def\yy{{\bf y}}
\def\nnn{{\bf n}}
\def\TT{{\bf T}}
\def\MM{{\bf M}}

\def\WW{{\cal W}}

\def\RR{{\cal R}}
\def\RRR{{\mathbb R}}
\def\CCC{{\mathbb C}}
\def\LL{{\cal L}}
\def\BB{{\cal B}}
\def\SS{{\cal S}}
\def\GG{{\cal G}}

\def\FF{{\cal F}}
\def\CC{{\cal C}}
\def\EE{{\cal E}}
\def\TTT{{\cal T}}
\def\ZZZ{{\mathbb Z}}
\def\AA{{\cal A}}

\def\L{\Lambda}
\def\l{\lambda}
\def\a{\alpha}
\def\n{\nu}
\def\b{\beta}
\def\m{\mu}
\def\d{\delta}
\def\g{\gamma}
\def\r{\rho}

\def\s{\sigma}
\def\e{\varepsilon}
\def\t{\vartheta}

\def\p{\pi}

\def\lb{\label}

\def\ul{\underline}

\def\const{\mbox{const}}

\renewcommand\Re{\operatorname{Re}}

\def\be{\begin{equation}}
\def\ee{\end{equation}}
\def\bea{\begin{eqnarray}}
\def\eea{\end{eqnarray}}

\def\nn{\nonumber}

\begin{document}

\title{Borel summability of $\varphi^{4}_{4}$ planar theory via multiscale analysis}

\author{Marcello Porta}
 \affiliation{Dipartimento di Fisica,
Universit\`a di Roma ``La Sapienza'', Piazzale Aldo Moro 5, 00185 Roma Italy}
\author{Sergio Simonella}
\affiliation{Dipartimento di Matematica,
Universit\`a di Roma ``La Sapienza'', Piazzale Aldo Moro 5, 00185 Roma Italy}

\begin{abstract}{We review the issue of Borel summability in the framework of multiscale analysis and renormalization group, by discussing a proof of Borel summability of the $\varphi^{4}_4$ massive euclidean planar theory; this result is not new, since it was obtained by Rivasseau and 't Hooft. However, the techniques that we use have already been proved effective in the analysis of various models of consended matter and field theory; therefore, we take the $\varphi^{4}_4$ planar theory as a toy model for future applications.}\end{abstract}

\maketitle

\section{Introduction}

The problem of giving a meaning to the {\em formal} perturbative series defining the scalar $\varphi^{4}_4$ theory, the simplest four dimensional interacting field theory, has been very debated (see \cite{GR} for a critical introduction to the problem) and it is still wide open, despite several triviality conjectures have been proposed since the work of Landau, \cite{L}. Here we focus on the {\em planar restriction} of the full perturbative series; that is, we consider only the graphs that can be drawn on a sheet of paper without ever crossing lines in points where no interacting vertices are present. This problem is much easier than the complete case, since the number of topological Feynman graphs contributing to a given order $n$ is much smaller than the original $n!$: in fact, in the planar theory this number is bounded by $(\const.)^{n}$, see \cite{K, BIPZ}. Still, the problem is far from being trivial, since the theory needs to be renormalized; this can be done using renormalization group, see \cite{GN1, GN2, G, P} for instance. 

It is well known that the full $\varphi^{4}_4$ with the ``wrong'' sign of the renormalized coupling constant, that is the one corresponding to an unstable self interaction potential, is perturbatively {\em asymptotically free}, in the sense that truncating the beta function to a finite order the {\em running coupling constant} describing the interaction of the fields at energy scale $\m$ flows to zero in the ultraviolet as $(\log\m)^{-1}$. This fact does not have any direct physical interpretation in the full $\varphi^{4}_4$, since the theory is not defined for the considered value of the renormalized coupling constant; moreover, the beta function itself is not defined, because of the factorial growth of the number of topological Feynman graphs in the order of the series.

However, these problems do not affect the planar theory, since it is only defined perturbatively and the number of graphs at a given order is far smaller than the original $n!$; therefore, one can hope in this case to exploit asymptotic freedom to rigorously construct the theory. This has been done independently by Rivasseau and 't Hooft using quite different methods, see \cite{R, R2, tH, tH2}; indeed, they proved that the renormalized perturbative series defining the Schwinger functions, which are the result of various resummations, are absolutely convergent. In particular, they proved that the result is the {\em Borel sum} of the perturbative series in the renormalized coupling constant. This last fact means in particular that the Schwinger functions can be expressed to an arbitrary accuracy starting from their perturbative series in the renormalized coupling constant, following a well defined prescription; moreover, the result is unique within a certain class of functions, the {\em Borel summable} ones. Clearly, this does not exclude the existence of other less regular solutions with the same formal perturbative expansion.

At the time of those works, besides the possibility of giving a mathematically rigorous meaning to a simple quantum field theory, the physical motivation of the study was that the $\varphi^{4}_4$ planar theory is formally equal to the limit $N\rightarrow \infty$ of a massive $SU(N)$ theory in four dimensions, with interaction $\lambda {\rm Tr}\, \varphi^{4}$ where $\varphi$ is an $N\times N$ matrix, see \cite{tH2, BIPZ}; in particular, in 't Hooft work the planar approximation was seen as a first step towards the more ambitious study of QCD with large number of colors.

In this paper we review the issue of Borel summability of the $\varphi^{4}_4$ planar theory using the rigorous renormalization group techniques introduced in \cite{GN1, GN2, G} (in \cite{GN2, G} the flow of the running coupling constants of the planar theory was heuristically discussed), which make possible a transparent proof of the ultraviolet stability of the massive euclidean $\varphi^{4}_4$ theory, through the so called ``$n!$ bounds''. 

One of the motivations of our work lies in the fact that very few proofs of Borel summability based on renormalization group methods are present in literature, \cite{FMRS, FMRS2}; moreover, we take the $\varphi^{4}_4$ planar theory as a first step towards the study of physically more interesting models, which can be analyzed by similar techniques. As mentioned before, the great gain that one has in the planar restriction of the full $\varphi^{4}_4$ theory is that the topological Feynman graphs of a given order $n$ are far less (their number is bounded as $(\const.)^{n}$, against the $n!$ of the full case). This is in a sense reminiscent of what happens in fermionic field theories, where it is possible to control the factorial growth of the number of Feynman graphs by exploiting the $-1$ arising in the anticommutation of the fields, showing that the $n$--th order of the series, which is given by $n!$ addends, reconstructs the determinant of an $n\times n$ matrix, which is estimated by $(\const.)^{n}$. For instance, we think that the methods described in this paper could be useful to prove Borel summability for the one dimensional Hubbard model, where one sector of the theory is asymptotically free, while to control the flow of the other running coupling constants one has to prove that the beta function is vanishing, \cite{BM}. This model has been rigorously constructed in \cite{M} using renormalization group methods similar to those used here, but a proof of Borel summability has not been given yet.

Informally, our main result can be stated as follows; we refer the reader to section \ref{sec3}, theorem \ref{th:1}, for a precise formulation.
\vskip.2cm
{\bf Main result.} {\em The Schwinger functions of the euclidean massive planar $\varphi^{4}_4$ theory are Borel summable in the renormalized coupling constant; in particular, they satisfy the hypothesis of the Nevanlinna -- Sokal theorem, \cite{S}, which are sufficient conditions for Borel summability.}
\vskip.2cm

Roughly speaking, our proof goes as follows. First, by choosing the renormalized coupling constant in a suitable complex domain, we prove that the flow equation defining recursively the running coupling constants at all energy scales admits a bounded solution which falls into the radius of convergence of the Schwinger functions, and verifies some special regularity properties; to do that we use a fixed point argument, similar to the one introduced by 't Hooft in \cite{tH}. Then, to conclude the check of the hypothesis of Nevanlinna--Sokal theorem on Borel summability, we show that it is possible to ``undo'' the resummation that allowed to write the Schwinger functions as power series in the running coupling constants so that the $n$--th order Taylor remainder in the renormalized coupling constant $\l$ can be bounded proportionally to $n!|\l|^{n+1}$ uniformly in the analyticity domain. To prove this second statement we rely in a crucial way on the {\em Gallavotti -- Nicol\` o tree representation} of the beta function; the ``undoing'' of the resummations, corresponding to rather involved analytical operations, is made clear by a graphical manipulation of these trees. This procedure is quite similar in spirit to what has been done by Rivasseau in \cite{R}.

Therefore, we feel that our proof lies halfway between those of Rivasseau and 't Hooft. As mentioned above, in 't Hooft approach, which is based on renormalization group ideas, the flow of the beta function is studied in a way analogous to the one we follow. However, instead of deriving bounds on the remainder of the resummed perturbative series, 't Hooft, see \cite{tH}, concludes the proof of Borel summability by checking the analyticity properties of the Borel transform using a totally independent argument, that we have not been able to rigorously reproduce in our framework.

For what concerns the comparison with Rivasseau's work, see \cite{R}, the main difference is that in his approach the beta function is not introduced: to construct the planar theory Rivasseau uses a ``minimal'' resummation procedure, involving only a certain class of Feynman graphs with four external legs, the {\em parquet} ones. This defines an asymptotically free ``running coupling constant'', and it turns out to be enough to prove the finiteness of the planar theory. To conclude, Rivasseau shows that the result of these operations is the Borel sum of the nonrenormalized series, by proving an $n!$ bound on the Taylor remainder; this bound is obtained undoing the resummation of the parquet subgraphs in a suitable way.

The paper is organized as follows. In section \ref{sec2} we define the model, we set the notations, we briefly review the ideas behind multiscale integration and we introduce the beta function and the flow of the running coupling constants; we refer the interested reader to \cite{GN1, GN2, G} for a detailed introduction to these techniques. In section \ref{sec3} we state our main result and we discuss the strategy of the proof. Finally, in section \ref{sec4} and in the appendices cited therein we prove the theorem.

\section{Renormalization group analysis}\lb{sec2}
\renewcommand{\theequation}{\ref{sec2}.\arabic{equation}}

\begin{figure}[htbp]
\centering
\includegraphics[width=1\textwidth]{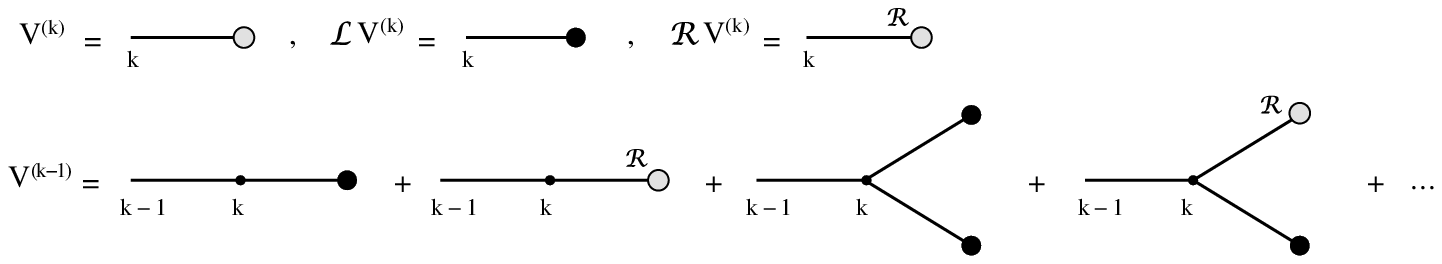}
\caption{Graphical interpretation of (\protect\ref{1.3.3}). The graphical equations for $\LL V^{(k-1)}$, $\RR V^{(k-1)}$ are obtained from the equation in the second line by putting an $\LL$, $\RR$ label, respectively, over the vertices on scale $k$.} 
\label{fig00}
\end{figure}

\begin{figure}[htbp]
\centering
\includegraphics[width=0.6\textwidth]{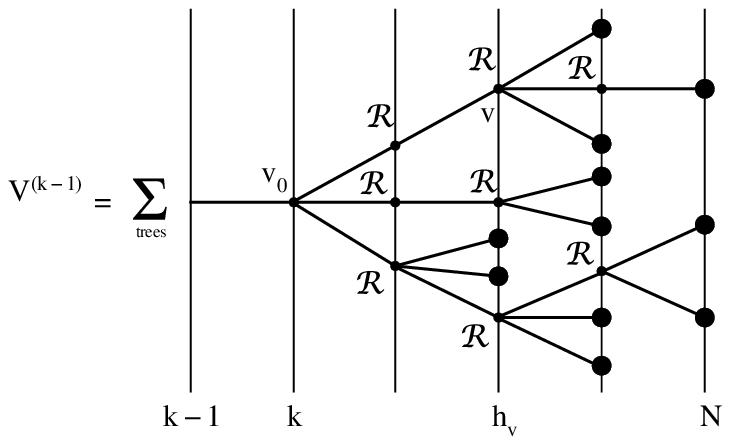}
\caption{The effective potential $V^{(h)}$
can be represented as a sum over {\em Gallavotti -- Nicol\`o} trees. The small
black dots will be called {\it vertices} of the tree. All the vertices except
the first ({\it i.e.} the one on scale $k$) have an $\RR$ label attached, which
 means that they correspond to the action of $\RR\EE^{T}_{h_v}$, while the
first represents $\EE^{T}_{k}$. The generic endpoint $e$, represented by a fat endpoint, corresponds to $\LL V^{(h_e -1)}$. The sum is over {\em distinct} trees; two trees are considered {\em identical} if it is possible to superpose them together with the labels appended to their vertices by stretching or shortening the branches.} \label{fig01}
\end{figure}

In this section we describe the iterative procedure that allows to express the Schwinger functions of the full $\varphi^{4}_4$ theory as power series order by order finite in the ultraviolet limit, graphically represented in terms of renormalized Feynman graphs; at the same time, we define the planar $\varphi^{4}_4$ theory by considering at each step only the planar graphs. Our discussion will be quite short; we refer the reader to \cite{G} for a detailed proof of the renormalizability of the $\varphi^{4}_4$ theory. In the following we shall denote by [A]-(B) the formula (B) of reference [A].
\vskip.1cm
{\it The full $\varphi^{4}_4$ theory.} Let $\varphi_{\xx}^{(\leq N)}$ be a massive gaussian free field with ultraviolet cut-off at length $\gamma^{-N}$, where $\gamma > 1 $ is a fixed scale parameter, and $\xx\in\L$ where $\L$ is a four--dimensional box of side size $L$ with periodic boundary conditions; for simplicity, we set to $1$ the value of the mass. We rewrite the field as:
\begin{equation}
\varphi^{(\leq N)}_{\xx} = \sum_{j=0}^{N}\varphi^{(j)}_{\xx}\;,\qquad \xx \in \L\;,\label{1.1}
\end{equation}
where $\{\varphi^{(j)}\}_{j=0}^N$ are independent gaussian fields with propagators
\be
C^{(j)}_{\xx, \yy} := \int \frac{d{\bf p}}{(2\pi)^4}\,\frac{f_j(\bf p)}{\pp^{2} + 1}e^{i{\bf p}\cdot (\bf x - \bf y)}\;,\quad f_{j}(\pp) := \left\{\begin{array}{cc} e^{-\pp^2/\gamma^{2j}} - e^{-\pp^{2}/\gamma^{2(j-1)}} & \mbox{if $j>0$}\\ e^{-\pp^2} & \mbox{if $j=0$} \end{array}\right.\;,\label{1.2}
\ee
and $\int \frac{d\pp}{(2\pi)^{4}}$ is a shorthand for $|\L|^{-1}\sum_{\pp = 2\pi\nnn/L}$ with $\nnn\in\ZZZ^{4}$; notice that
\be
\lim_{N\rightarrow +\infty}\sum_{j=0}^{N}f_{j}(\pp) = 1\;.\label{1.2b}
\ee
The generating functional of the Schwinger functions of the $\varphi^{4}_4$ theory is given by:
\be
e^{\WW_{N}(\zeta f)} := \int \exp\left(\zeta\int d\xx\, \varphi^{(\leq N)}_{\xx}f_{\xx}\right)e^{V^{(N)}(\varphi^{(\leq N)})} P(d\varphi^{(\leq N)})\;,\label{1.2c}
\ee
where: $f_{\xx}$ is a Schwartz test function, $\zeta\in \RRR$, $P(d\varphi^{(\leq N)}):= \prod_{j=0}^{N}P(d\varphi^{(j)})$ with $P(d\varphi^{(j)})$ the gaussian distribution of the field $\varphi^{(j)}$ with covariance given by (\ref{1.2}), and the {\em interaction} $V^{(N)}$ is defined as
\be
V^{(N)}(\varphi^{(\leq N)}) := \int_{\L} d\xx\; \left(\l_N:(\varphi^{(\leq N)}_{\xx})^4: + \a_N:(\partial\varphi^{(\leq N)}_{\xx})^{2}: + \m_N:(\varphi^{(\leq N)}_{\xx})^2: + \n_N\right)\;, \label{1.3}
\ee
where $\l_N,\a_N,\m_N,\n_N$ are called {\em bare coupling constants}; notice that in our convention the ``wrong'' sign of $\l_N$ is the positive one. The generic $q$-point Schwinger function of the full $\varphi^{4}_4$ theory is obtained deriving the generating functional $q$ times with respect to $\zeta$ and setting $\zeta=0$. Now, let $\WW_{N}(\zeta f) =: \WW_{N}^{p}(\zeta f) + \WW_{N}^{np}(\zeta f)$, where $\WW_{N}^{p},\WW_{N}^{np}$ are respectively the {\em planar}/{\em non planar} part of $\WW_N$ to be defined recursively in the following; the $q$-point Schwinger function of the planar theory is defined as:
\be
S^{T}_{(N)}(f;q) := \frac{\partial^q}{\partial \zeta ^{q}} \WW^{p}_{N}(\zeta f)\Big|_{\zeta=0}\;.\lb{1.3ba}
\ee
We shall denote by $S^{T}(f;q)$ the limit for $N\rightarrow+\infty$ of (\ref{1.3ba}).
\vskip.1cm
{\it Multiscale analysis.} As explained in \cite{G}, we can try to evaluate (\ref{1.2c}) by proceeding in an iterative fashion, integrating the independent fields $\varphi^{(j)}$ starting from the ultraviolet scale $j=N$ going down to the infrared scale $j=0$. This iterative integration gives rise to an expansion in Feynman graphs; the restriction to the planar theory will be enforced by considering at each integration step only the planar ones. For simplicity, in what follows we shall explicitly discuss only the case $f=0$, which corresponds to the integration of the ``partition function''; the case $f\neq 0$ is a straightforward extension of our argument, and it will be discussed later. After the integration of $\varphi^{(N)},\varphi^{(N-1)},\ldots,\varphi^{(k+1)}$ we rewrite the integral (\ref{1.2c}) as
\be
e^{\WW_{N}(0)} = \int e^{V^{(k)}(\varphi^{(\leq k)})} P(d\varphi^{(\leq k)}) = \int e^{V^{(k)}_{p}(\varphi^{(\leq k)}) + V^{(k)}_{np}(\varphi^{(\leq k)})}P(d\varphi^{(\leq k)})\;,\label{1.3b}
\ee
where: $P(d\varphi^{(\leq k)}) := \prod_{j=0}^{k}P(d\varphi^{(j)})$, the field $\varphi^{(\leq k)} = \sum_{j=0}^{k}\varphi^{(j)}$ has a propagator given by, in momentum space,
\be
C^{(\leq k)}_{\pp} := \sum_{j=0}^{k}C^{(j)}_{\pp}\;,\qquad C^{(j)}_{\pp} := \frac{f_j(\pp)}{\pp^2 + 1}\;,
\ee
and the {\em effective potential} $V^{(k)}$ together with its planar/non planar parts $V^{(k)}_{p}$, $V^{(k)}_{np}$ will be defined recursively. At the beginning, $V^{(N)}(\varphi^{(\leq N)}) = V^{(N)}_{p}(\varphi^{(\leq N)})$; on scale $k$ we will show that, if $\# = ``p", ``np"$:
\be
V^{(k)}_{\#}(\varphi^{(\leq k)}) = \sum_{m\geq 0}\int \frac{d\pp_1}{(2\pi)^4}\ldots \frac{d\pp_m}{(2\pi)^4} V^{(k)}_{\#}(\pp_1,\ldots \pp_m;m):\prod_{i=1}^{m}\varphi^{(\leq k)}_{\pp_i}:\delta\big(\sum_{i}\pp_i\big)\;,\label{1.3cc2}
\ee
where $V^{(k)}_{\#}(\pp_1,\ldots,\pp_m;m)$ are suitable coefficients to be recursively defined, and the product with $m=0$ is interpreted as $1$. Let us perfom the single scale integration. First, we split $V^{(k)}$ as $\LL V^{(k)} + \RR V^{(k)}$, where $\RR = 1-\LL$ and $\LL$, the {\em localization operator}, is a linear operator acting on functions of the form (\ref{1.3cc2}), defined by its action on the kernels $V^{(k)}_{\#}(\pp_1,\ldots,\pp_m;m)$ in the following way (with a slight abuse of notation, due to the presence of the delta function in (\ref{1.3cc2}) we only write the independent values of the momenta in the arguments of the kernels):
\bea
&&\LL V^{(k)}_{\#}(\pp_1,\pp_2,\pp_3;4) := V^{(k)}_{\#}({\bf 0},{\bf 0},{\bf 0};4)\;,\label{1.3d}\\
&&\LL V^{(k)}_{\#}(\pp;2) := V_{\#}^{(k)}({\bf 0};2) + \pp\partial_\pp V^{(k)}_{\#}({\bf 0};2) + \frac{1}{2}p_{i}p_{j}\partial_{p_i}\partial_{p_j}V^{(k)}_{\#}({\bf 0};2)\;,\nn
\eea 
and $\LL V^{(k)}_{\#}(\pp_1,\ldots,\pp_m;m) = 0$ otherwise. By symmetry, it follows that
\bea
&&\partial_{p_i}V^{(k)}_{\#}({\bf 0};2) = 0\;,\qquad \partial_{p_i}\partial_{p_j}V^{(k)}_{\#}({\bf 0};2) = 0\quad \mbox{for $i\neq j$}\;,\nn\\
&&\partial_{p_{i}p_{i}}V^{(k)}_{\#}({\bf 0};2) = \partial_{p_{j}p_{j}}V^{(k)}_{\#}({\bf 0};2)\quad \mbox{for all $i,j$}\;;\label{1.3.1}
\eea
finally, we define the {\em running coupling constants of the planar theory} on scale $k$ as:
\be
\l_k := V^{(k)}_{p}({\bf 0},{\bf 0},{\bf 0};4)\;,\quad \g^{2k}\m_k := V^{(k)}_{p}({\bf 0};2)\;,\quad \a_k := \frac{1}{2}\partial_{p_1 p_1} V^{(k)}_{p}({\bf 0};2)\;,\quad \g^{4k}\n_k:= V^{(k)}_{p}(0)\,;\label{1.3.2}
\ee
the corresponding objects in the full theory are obtained by replacing the $V_{p}^{(k)}$ in (\ref{1.3.2}) with $V^{(k)}$. Therefore, setting $\varphi^{(\leq k)} =: \varphi^{(\leq k-1)} + \varphi^{(k)}$, we can rewrite (\ref{1.3b}) with $k$ replaced by $k-1$, and $V^{(k-1)}$ given by
\be
V^{(k-1)}(\varphi^{(\leq k-1)}) = \log \int P(d\varphi^{(k)}) e^{V^{(k)}(\varphi^{(\leq k-1)} + \varphi^{(k)})} := \sum_{n\geq 0}\frac{1}{n!}\EE^{T}_k(V^{(k)}(\varphi^{(\leq k)});n)\;,\label{1.3.3}
\ee
where $\EE^{T}_k$ is called {\em truncated expectation on scale $k$}, and it is defined as:
\be
\EE^{T}_h(X(\varphi^{(h)});n):= \frac{\partial^{n}}{\partial \zeta^{n}}\log \int P(d\varphi^{(h)})e^{\zeta X(\varphi^{(h)})}\Big|_{\zeta=0}\;.\label{1.3.4}
\ee 
It is convenient to define also $V^{(-1)}$; for this purpose one thinks $\varphi^{(\leq N)}$ as being given by, see formula \cite{G}-(6.9),
\be
\varphi^{(\leq N)} = \varphi^{(-1)} + \varphi^{(0)} + \ldots + \varphi^{(N)}\;,\label{1.3bb}
\ee
where the field $\varphi^{(-1)}$ is distributed independently relative to the other $\varphi^{(j)}$, $j\geq 0$, and it has its own covariance $C^{(-1)}_{\xx,\yy}$ which needs not to be specified (because it will eventually be taken to be identically zero whenever it appears in some interesting formulas). The introduction of $V^{(-1)}$ allows to treat the case $k=0$ on the same grounds as the cases $k>0$.
\vskip.1cm
{\it Tree expansion and Feynman graphs.} The iterative integration described above leads to a representation of the effective potential on scale $k-1$ as a power series in the running coupling constants $\l_h,\a_h,\m_h,\n_h$ with $h\geq k$, where the coefficients of the series can be represented in terms of connected Feynman graphs, as briefly explained in the following. The key formula which we start from is (\ref{1.3.3}); iterating this formula as suggested by figure \ref{fig00}, we end up with a representation of the effective potentials in terms of a sum over {\em Gallavotti -- Nicol\` o} trees, \cite{GN1, GN2, G}, see figure \ref{fig01}:
\bea
&&V^{(k-1)}(\varphi^{(\leq h)}) = \sum_{n\geq 1}\sum_{\g\in \TTT_{k-1,n}}V^{(k-1)}(\g)\;,\label{1.3.5}\\
&& V^{(k-1)}(\g) = \sum_{m\geq 0}\int \frac{d\pp_1}{(2\pi)^4}\ldots\frac{d\pp_m}{(2\pi)^4}V^{(k-1)}(\pp_1,\ldots ,\pp_{m};\g,m):\prod_{i=1}^{m}\varphi^{(\leq k-1)}_{\pp_i}:\d\big(\sum_{i}\pp_i\big)\;,\nn
\eea
where $\TTT_{k-1,n}$ is the set of trees with {\em root} $r$ on scale $h_r = k-1$ and $n$ endpoints, with value $V^{(k-1)}(\g)$. The trees involved in the sum are {\em distinct}; two trees are considered {\em identical} if it is possible to superpose them together with the labels appended to their vertices by stretching or shortening the branches. 
Proceeding in a way analogous to \cite{G}, section XVI and appendix C, it follows that the kernels $V^{(k)}(\pp_1\;,\ldots\;,\pp_{m};\g,m)$ satisfy the following recursion relation:
\bea
&&V^{(k-1)}(\pp_1,\ldots,\pp_m;\g,m) = \sum_{m_1,\ldots,m_s}\frac{1}{s!}\int \Big[ \prod_{j=1}^{s}\widetilde V^{(k)}(\pp_1,\ldots ,\pp_{m_j}; \g_j,m_j) \Big]\cdot \nn\\&&\hskip5cm \cdot \sum_{\pi\in \GG_m}\sum_{\t \subset \pi\atop connected}\Big[ \prod_{\l\in \t}C^{(k)}_{\pp(\l)} \Big]\cdot\Big[ \prod_{\l\in \pi/\t}C^{(\leq k-1)}_{\pp(\l)} \Big]\;,\label{1.3.6}
\eea
where $\g_1,\ldots,\g_s$ are the $s$ subtrees of $\g$ with root corresponding to the first nontrivial vertex of $\g$, $\widetilde V^{(k)}(\pp_1,\ldots ,\pp_{m_j}; \g_j,m_j)$ is equal to $\RR V^{(k)}(\pp_1,\ldots ,\pp_{m_j}; \g_j,m_j)$ if $\g_j$ is nontrivial and to $\LL V^{(k)}(\pp_1,\ldots ,\pp_{m_j}; m_j)$ otherwise, $\GG_m$ is a suitable set of Feynman graphs defined below, and the integral is over their loop momenta; this relation is a consequence of the rules  of evaluation of the truncated expectations of Wick monomials, see \cite{G} appendix C. Formula (\ref{1.3.6}) is iterated by replacing each $\widetilde V^{(k)}(\pp_1,\ldots ,\pp_{m_j}; \g_j,m_j)$ corresponding to nontrivial $\g_j$'s with (\ref{1.3.6}) with $k-1$ replaced by $k$. Analogously, the {\it planar part} of the effective potential is defined as:
\bea
&&V^{(k-1)}_{p}(\pp_1,\ldots,\pp_m;\g,m) = \sum_{m_1,\ldots,m_s}\frac{1}{s!}\int \Big[ \prod_{j=1}^{s}\widetilde V^{(k)}_{p}(\pp_1,\ldots ,\pp_{m_j}; \g_j,m_j) \Big]\cdot \nn\\&&\hskip5cm \cdot \sum_{\pi\in \GG_m\atop \pi\,planar}\sum_{\t \subset \pi\atop connected}\Big[ \prod_{\l\in \t}C^{(k)}_{\pp(\l)} \Big]\cdot\Big[ \prod_{\l\in \pi/\t}C^{(\leq k-1)}_{\pp(\l)} \Big]\;.\label{1.3.6.2}
\eea
Represent a generic Wick monomial $M_j$ containing the product of $m_j$ fields as a point or as a cluster with $m_j$ emerging lines, depending on whether the corresponding $\g_j$ is trivial or not; we shall consider the points as (trivial) clusters, too. Given the Wick monomials $M_{1},\ldots , M_{s}$ the symbol $\GG_m$ denotes the set of connected graphs that can be made joining pairwise some of the lines associated with the clusters $M_{1},\ldots , M_{s}$ in such a way that: (i) two lines emerging from the same cluster cannot be contracted together, (ii) there should be enough lines so that looking the clusters as points the resulting graph is connected, (iii) after the contraction there should be still $m$ uncontracted lines, representing the Wick monomial $M$. The resulting graph is enclosed in a new cluster, labelled by $k$. Furthermore, the condition $\t\subset \p$ with the subscript ``connected'' means that the subgraph $\t$ still keeps the connection between the boxes. We graphically represent the propagators $C^{(k)}$ by a solid line, while $C^{(\leq k-1)}$ correspond to wavy lines. Finally, the restriction to planarity means that we discard all the graphs that show lines crossing in points were no interacting vertices are present. We refer the reader to \cite{G}, section XVI, for a more extensive discussion and for examples.

\begin{figure}[htbp]
\centering
\includegraphics[width=0.4\textwidth]{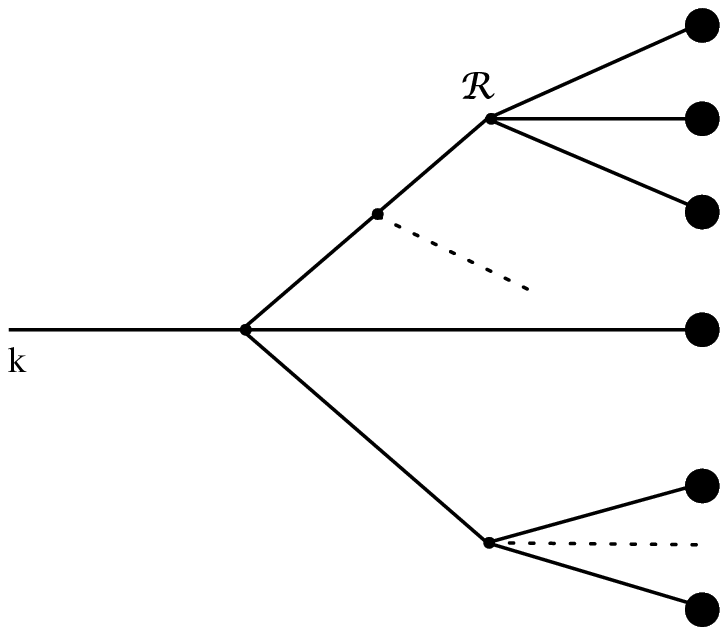}
\caption{A generic tree belonging to $\TTT_{k,6,2}$.} \label{fig:1b}
\end{figure}

Clearly, the iteration stops when only trivial subtrees appear in (\ref{1.3.6}), (\ref{1.3.6.2}); at this point, the resulting graph looks like an ``usual'' one, but enclosed in a hierarchical cluster structure, where each cluster has a scale label; and given two clusters $G_{v}$, $G_{v'}$ then $G_{v}\subset G_{v'}$ if and only if $h_{v}>h_{v'}$. After the iteration, the effective potential on scale $k$ is expressed as a power series in the running coupling constants $\l_h,\a_h,\m_h,\n_h$ with $h>k$. From the analysis of \cite{GN1, GN2, G} it follows that the contribution of a given tree $\g\in\TTT_{k,n}$ to a kernel of the planar theory can be bounded in the following way, setting $\d := \max_{h}\{|\l_h|,|\a_h|,|\m_h|,|\n_h|\}$, for some positive $C_{m},\r$:
\be
\big| V_p^{(k)}(\pp_1,\ldots, \pp_m; \g, m) \big| \leq C_m(\const.)^{n}\d^{n}\g^{k(4 - m)}\prod_{\substack{v>r \atop v\,not\,e.p.}}\g^{-\r(h_{v} - h_{v'})}\;,\label{1.3.7.0}
\ee
where the product runs over the vertices of the tree $\g$ and $v'$ is the vertex immediately preceding $v$; since the number of distinct trees is bounded as $(\const.)^{n}$ it follows that, see \cite{G} section XIX:
\bea
\sum_{\g\in \TTT_{k,n}} |V^{(k)}_{p}(\pp_1,\ldots,\pp_m;\g,m)| &\leq& C_m C^{n}\d^{n}\g^{k(4 - m)}\;\;,\label{1.3.7}
\eea
which means that the planar part of the effective potential can be expressed as a convergent power series in the running coupling constants, provided their absolute values are small enough. This is not the case in the full theory; in the analogous of (\ref{1.3.7.0}), due to the combinatorics of the Feynman graphs, one has to take into account an extra $n!$ factor. Formula (\ref{1.3.7.0}) implies in particular the so called {\em short memory} property of the Gallavotti -- Nicol\` o trees, which states that if two scales of a given tree are constrained to have fixed values, say $h,k$ with $h<k$, then the bound on the sum over all the remaining scales is improved by a factor $\g^{-(\r/2)(k - h)}$ with respect to (\ref{1.3.7}); in other words, long trees are exponentially suppressed.
\vskip.1cm
{\it The expansion of the Schwinger functions.} The generating functional of the Schwinger functions can be evaluated repeating a procedure completely analogous to the one described for the effective potentials; after the integration of the scales $N,N-1,\ldots,k+1$ it turns out that:
\be
e^{\WW_{N}(\zeta f)} = \int P(d\varphi^{(\leq k)})e^{S^{(k)}(\varphi^{(\leq k)};\zeta f)} = \int P(d\varphi^{(\leq k)})e^{S_p^{(k)}(\varphi^{(\leq k)};\zeta f) + S_{np}^{(k)}(\varphi^{(\leq k)};\zeta f)}\;,\label{1.3.7b}
\ee
where the effective potentials $S_{\#}^{(k)}(\varphi^{(\leq k)};\zeta f)$ have the form:
\be
S_{\#}^{(k)}(\varphi^{(\leq k)};\zeta f) =  \sum_{m\geq 0\atop t\geq 0}\int \frac{d\pp_1}{(2\pi)^4}\ldots \frac{d\pp_{m+t}}{(2\pi)^4}\, S^{(k)}_{\#}(\pp_1,\ldots \pp_{m+t};m,t)\cdot\Big[:\prod_{i=1}^{m}\varphi^{(\leq k)}_{\pp_i}:\prod_{j=1}^{t}\zeta f_{\pp_{j}}\Big]\;,\quad \label{1.3.7c}
\ee
and can be represented as sums over trees very similar to the ones introduced for the effective potentials, up to the following differences, see \cite{GN1, GN2} section 7.5: (i) special vertices may appear, from which dotted lines representing the ``external fields'' $\zeta f$ emerge (that do not contribute to the total number of endpoints), and (ii) no $\RR$ operation is defined on the path from a given dotted line to the root. We call $\TTT_{k,n,t}$ the set of such trees having root scale $k$, $n$ endpoints and $t$ dotted lines. See figure \ref{fig:1b} for an example.

\begin{figure}[htbp]
\centering
\includegraphics[width=0.8\textwidth]{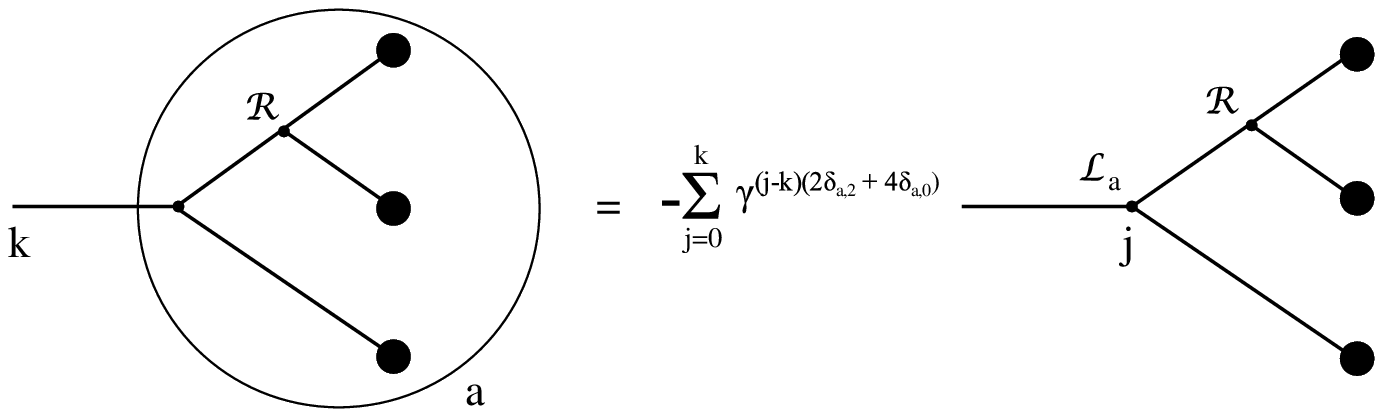}
\caption{Example of framed tree.} \label{fig02}
\end{figure}

Setting
\be
S^{(k)}_{\#}(\pp_1,\ldots \pp_{m+t};m,t) = \sum_{n\geq 1}\sum_{\g\in\TTT_{k,n,t}}S_{\#}^{(k)}(\pp_1,\ldots,\pp_{m+t};\g,m,t)\;,
\ee
the planar parts of the kernels of the effective potentials are related by the following recursive equation:
\bea
&&S_{p}^{(k-1)}(\pp_1,\ldots,\pp_{m+t};\g,m,t) = \sum_{m_1,\ldots,m_s\atop t_{1},\ldots , t_{s}}\frac{1}{s!}\int \Big[ \prod_{j=1}^{s}\widetilde S^{(k)}_{p}(\pp_1,\ldots ,\pp_{m_j+t_{j}}; \g_j,m_j,t_{j}) \Big]\cdot \nn\\&&\hskip5cm \cdot \sum_{\pi\in \GG_m\atop \pi\,planar}\sum_{\t \subset \pi\atop connected}\Big[ \prod_{\l\in \t}C^{(k)}_{\pp(\l)} \Big]\cdot\Big[ \prod_{\l\in \pi/\t}C^{(\leq k-1)}_{\pp(\l)} \Big]\;,\label{1.3.7.3}
\eea
where: $\g_1,\ldots , \g_s$ are the $s$ subtrees of $\g$ with root coinciding with the first vertex of $\g$ following the root; if $\g_j$ is trivial and corresponds to a dotted line then $\widetilde S^{(k)}_{p}(\pp_1,\ldots ,\pp_{m_j+t_j}; \g_j,m_j,t_j) = \d_{m_j,1}\d_{t_j,1}$, while if it corresponds to a solid line $\widetilde S^{(k)}_{p}(\pp_1,\ldots ,\pp_{m_j+t_j}; \g_j,m_j,t_j) = \d_{t_j,0}\LL V^{(k)}(\pp_1,\ldots ,\pp_{m_j}; m_j)$; if $\g_j$ is a nontrivial subtree with $t_{j}>0$ then $\widetilde S^{(k)}_{p}(\pp_1,\ldots ,\pp_{m_j+t_j}; \g_j,m_j,t_j) = S^{(k)}_{p}(\pp_1,\ldots ,\pp_{m_j+t_j}; \g_j,m_j,t_j)$, while if $\g_j$ is nontrivial and $t_j = 0$ then $\widetilde S^{(k)}_{p}(\pp_1,\ldots ,\pp_{m_j}; \g_j,m_j,0) = \RR V^{(k)}_{p}(\pp_1,\ldots ,\pp_{m_j}; \g_j,m_j)$, with $\RR = 1 -\LL$ defined as in (\ref{1.3d}). Clearly, $m_{1},\ldots, m_{s}$ and $t_{1},\ldots , t_{s}$ are subject to the constraints $\sum_{j} m_{j} = m$, $\sum_{j} t_{j} = t$. Formula (\ref{1.3.7.3}) is iterated by replacing each $S^{(k)}_{p}(\pp_1,\ldots ,\pp_{m_j}; \g_j,m_j,t_j)$ corresponding to any nontrivial $\g_j$ with $t_j>0$. Therefore, the generic planar Schwinger function $S^{T}_{(N)}(f;q)$ can be written as:
\bea
&&S^{T}_{(N)}(f;q) = \sum_{n\geq 1}\sum_{\g\in\TTT_{-1,n,q}}S_{p}(\g)\;,\nn\\
&&S_{p}(\g) := \int \frac{d\pp_1}{(2\pi)^4}\ldots \frac{d\pp_q}{(2\pi)^4}\, f_{\pp_1}\ldots f_{\pp_q} S^{(-1)}_{p}(\pp_1,\ldots, \pp_{q};\g,0,q)\;,\quad \g\in \TTT_{-1,n,q}\;, \label{1.3.7.2}
\eea
where $S^{(-1)}_p$ is given by (\ref{1.3.7.3}) with $k=0$. Finally, from the theory of \cite{GN1}, see section 7.5, it follows that
\be
\sum_{\g\in\TTT_{-1,n,q}}|S_{p}(\g)|\leq \|f\|_1^{q}C_qC^{n}\d^{n}\;,\label{1.3.8}
\ee
which implies that in the planar theory the Schwinger functions can be expressed as {\it absolutely convergent} power series in the running coupling constants, provided their absolute values are small enough. As it is well known, this is not the case in the full theory, since the bound (\ref{1.3.8}) has to be multiplied by $n!$; see \cite{DR, GN1, GN2, G}.
\vskip.1cm
{\it The beta function and its tree expansion.} From now on we shall focus only on the planar theory. The running coupling constants obey to recursive equations induced by iterative integration; it follows that, setting $v^{(4)}_k := \l_k$, $v^{(2')}_{k} := \a_k$, $v^{(2)}_{k} = \m_k$, $v^{(0)}_{k} := \n_k$:
\be
v^{(a)}_{k}=\gamma^{-2\delta_{a,2}-4\delta_{a,0}}v^{(a)}_{k-1}-\left(\mathcal{B}\underline{\underline{v}}\right)^{(a)}_{k}\;,\qquad k\geq 0\;, \label{1.3c} 
\ee
where the operator $\mathcal{B}$, the {\em beta function} of the theory, has the form, see formula \cite{G}-(9.15):
\be
\left(\mathcal{B}\underline{\underline{v}}\right)^{(a)}_{k}:= \sum_{r=2}^{\infty}\sum_{h_{1},...,h_{r}\geq
k}^N\sum_{a_{1},...,a_{r}}\beta^{(a)}_{a_{1},...,a_{r}}(k;h_{1},...,h_{r})\prod_{i=1}^{r}v^{(a_{i})}_{h_{i}}\;.\label{1.4}
\ee 
The quantities $\{v_{-1}^{(a)}\}$ are called the {\em renormalized coupling constants}. As the iterative procedure described before suggests, the beta function $\left(\mathcal{B}\underline{\underline{v}}\right)^{(a)}_{k}$ can be represented as a sum over trees; the only difference with respect to the trees which have been introduced previously is that we attach an $\LL_{a}$ over the first vertex, where $\LL_{a}$ is defined in the following way: $\LL_{a}V_{p}^{(k)}(\pp_1,\pp_2,\pp_3;4) := V_{p}^{(k)}({\bf 0},{\bf 0},{\bf 0};4)$ if $a=4$ and zero otherwise, $\LL_{a}V_{p}^{(k)}(\pp;2):= \g^{-2k}V^{(k)}_{p}({\bf 0};2)$ if $a=2$ and zero otherwise, $\LL_{a}V^{(k)}_{p}(\pp;2):= (1/2)\partial_{p_{1}p_{1}}V^{(k)}_{p}({\bf 0};2)$ if $a=2'$ and zero otherwise and finally $\LL_{a}V^{(k)}_{p}(0):= \g^{-4k}V^{(k)}_{p}(0)$ if $a=0$ and zero otherwise. From the theory of \cite{G} it follows that in the planar theory:
\be
\sum_{h_{1},\ldots,h_{r}\geq k}\big|\beta^{(a)}_{a_{1},\ldots,a_{r}}(k;h_{1},\ldots,h_{r})\big|\leq (\const.)^{r}\;,\label{1.4.0}
\ee
which means that the beta function is defined as an {\em absolutely convergent} power series provided the absolute values of the running coupling constants are small enough; this is not the case in the full theory, since in that case the bound (\ref{1.4.0}) has to be multiplied by $r!$.
\vskip.1cm
{\bf Remarks.}
\begin{enumerate}
\item From the representation of the coefficients of the beta function in terms of Feynman graphs, induced by the iterative integration previously described (see also \cite{G} sections IX, XVI -- XIX), it follows that for $k>0$, calling $\bar r$ the number of indexes $i$ such that $a_i = 4$ (corresponding to the number of vertices with four external lines),
\bea
&&\beta^{(4)}_{a_{1},...,a_{r}}(k;h_{1},...,h_{r})=0 \qquad\mbox{unless $\bar r\geq 2$}\;,\label{1.4a}\\
&&\beta^{(2')}_{a_{1},...,a_{r}}(k;h_{1},...,h_{r})=0 \qquad\mbox{unless $\bar r\geq 2$}\;,\label{1.4b}\\
&&\beta^{(2)}_{a_{1},...,a_{r}}(k;h_{1},...,h_{r})=0 \qquad\mbox{unless $\bar r\geq 1$}\;.\label{1.4c}
\eea
These properties can be understood in the following way. The graphs contributing to (\ref{1.4a}) -- (\ref{1.4c}) are all computed at vanishing external momenta, and the momenta flowing on the propagators must have absolute values bigger than $0$; in fact, the quantity $\left(\mathcal{B}\underline{\underline{v}}\right)^{(a)}_{k}$ arise from the integration of the fields $\varphi^{(h)}$ with $h\geq k$, which if $k>0$ have support for momenta $\pp$ such that $|\pp|>0$. Then, to see property (\ref{1.4a}), simply try to draw on a sheet of paper any graph with four external lines evaluated at vanishing external momenta; as the reader may check, the condition $\bar r<2$ is not compatible with the fact that the momenta flowing on the propagators have absolute values $>0$. Property (\ref{1.4c}) can be seen in an analogous way. To understand (\ref{1.4b}), notice that the graphs contributing to $\beta^{(2')}_{a_{1},\ldots,a_{r}}(k;h_{1},\ldots,h_{r})$ have two external lines, and are derived twice with respect to the external momentum; then, proceed as for (\ref{1.4c}), and notice that the only two legged graphs with $\bar r = 1$ compatible with the request on the modulus of the inner momenta are ``tadpole'' graphs, which do not depend on the value of the external momentum; therefore, their derivatives are vanishing.
\item Note that the flow of $\n_{k}$ is {\em decoupled} from the others, since $\n_{k}$ does not
appear in the recursive equations defining $\lambda_{k}$, $\alpha_{k}$, $\mu_{k}$ (it is graphically represented by a vertex with no external lines); moreover, the sequence $\n_{-1},\ldots ,\n_{N}$ solves the following equation:
\begin{equation}
\n_{k} = \gamma^{-4}\n_{k-1} -
\left(\mathcal{B}\ul{\ul{v}}\right)^{(0)}_{k}\;,\label{1.5}
\end{equation}
which implies
\begin{equation}
\n_{k} = \gamma^{-4(k+1)}\n_{-1} -
\sum_{j=0}^{k}\gamma^{4(j-k)}\left(\mathcal{B}\ul{\ul{v}}\right)^{(0)}_{j}\;,\label{1.6}
\end{equation}
where $(\mathcal{B}\ul{\ul{v}})^{(0)}_{j}$ is analytic in its arguments for
$\max_{k}\{|\lambda_{k}|, |\alpha_{k}|, |\mu_{k}|\}$ small enough.
For these reasons, in what follows we shall focus only on the flows of $\lambda_{k}$, $\mu_{k}$,
$\alpha_{k}$.
\end{enumerate}
We can rewrite equation (\ref{1.3c}) as:
\be
v^{(a)}_{k} = \g^{-(k+1)(2\delta_{a,2}+4\delta_{a,0})}v^{(a)}_{-1} - \sum_{j=0}^{k}\g^{(j-k)(2\delta_{a,2}+4\delta_{a,0})}\left(\mathcal{B}\underline{\underline{v}}\right)^{(a)}_{j}\;,\label{1.4.1}
\ee
and this equation can be iterated in order to obtain the formal power series of $v^{(a)}_{k}$ in the renormalized coupling constants. Again, equation (\ref{1.4.1}) can be represented graphically. The second term in (\ref{1.4.1}) corresponds to the sum of all the possible trees with root scale $k$ enclosed in a {\em frame} labelled by a type label $a$. The correspondence between the framed trees and the trees discussed after (\ref{1.4}) is made explicit by the example in figure \ref{fig02}.

\begin{figure}[htbp]
\centering
\includegraphics[width=0.9\textwidth]{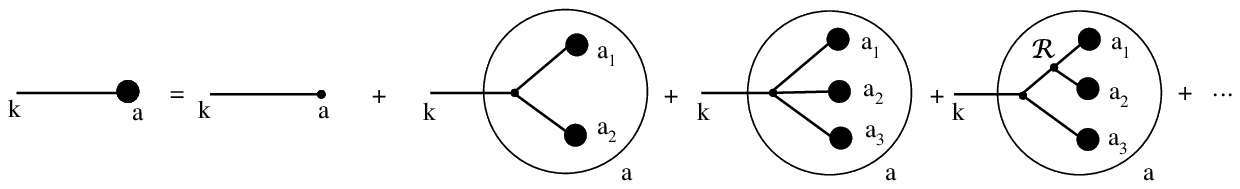}
\caption{Graphical interpretation of formula (\protect\ref{1.4.1}); a sum over the $a_{i}$'s is understood.}\label{fig03}
\end{figure}

\begin{figure}[htbp]
\centering
\includegraphics[width=1\textwidth]{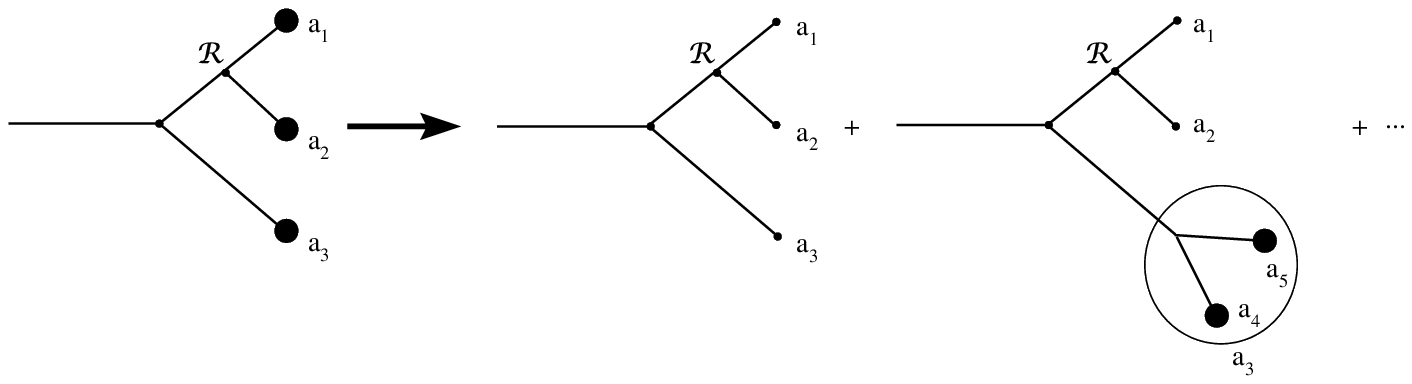}
\caption{Graphical interpretation of the iteration of equation (\protect\ref{1.4.1}).} \label{fig04}
\end{figure}

In general, the fat endpoint $e$ labelled by $a_{e}$ and attached to a vertex on scale $h_{e}-1$ corresponds to the running coupling constant $v^{(a)}_{h_{e}-1}$, while the first term in (\ref{1.4.1}) is represented as a trivial tree with a thin endpoint labelled by $a$ and root scale $k$. See figure (\ref{fig03}) for a graphical representation of (\ref{1.4.1}).
The iteration of (\ref{1.4.1}) produces trees showing thin endpoints, and in general more than one frame; see figure (\ref{fig04}) for a picture of the situation. 
Therefore, the $n$-th order contribution in the renormalized coupling constant to $v^{(a)}_{k}$ is defined graphically as the sum of all the possible framed trees with root scale $k$ enclosed in a frame labelled by $a$, with $n$ thin endpoints, and where the generic vertex $v$ has an $\RR$ label attached otherwise the corresponding subtree is enclosed in a frame. We stress that trees with different type labels attached to their frames and endpoints are considered {\em different}. The same graphical procedure allows to find the perturbative expansion of the Schwinger functions (or equivalently of the effective potentials) in the renormalized coupling constants, starting from their definition as trees with only ``fat'' endpoints. 
\vskip.1cm
{\bf Remark.} Given a generic framed tree showing any number of inner frames, we define the {\em maximally pruned framed tree} as the tree obtained by replacing the maximal inner frames ({\em i.e.} the ones enclosed only by the outermost frame) with fat endpoints of the corresponding type; by properties (\ref{1.4a})--(\ref{1.4c}) the sum over the scale of the first vertex of a framed tree, see figure \ref{fig02}, involves only the term with $j=0$ if:
\begin{itemize}
\item the type label of the frame is $2$ and the maximally pruned framed tree has no endpoints of type $4$;
\item the type label of the frame is $2'$ or $4$ and the maximally pruned framed tree has at most one endpoint of type $4$.
\end{itemize}
We shall say that a frame is {\em trivial} if the enclosed tree verifies one of the above properties; all the other frames will be called {\em nontrivial}.
\vskip.2cm
Call $\widetilde \TTT_{-1,m,q}$ the set of trees with root scale $-1$, any number of frames,  $m$ endpoints fat or thin, and $q$ dotted lines; given a generic tree $\g\in\widetilde\TTT_{-1,m,q}$ we call $n_{2',4}(\g)$ the number of nontrivial frames (see previous remark) labelled by $a=2',4$ and we denote by $m_a(\g)$ the number of endpoints of type $a$. In the planar theory the following remarkable result is true.

\begin{figure}[htbp]
\centering
\includegraphics[width=0.9\textwidth]{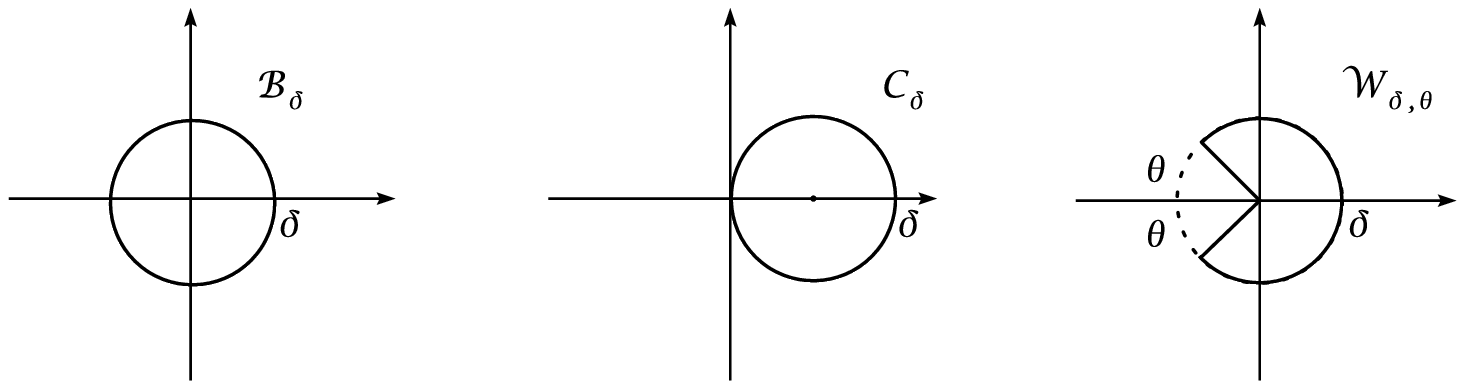}
\caption{The domains $\BB_\d$, $\CC_\d$, $\WW_{\d,\t}$.} \label{fig0}
\end{figure}

\begin{thm*}[{\bf $n!$ bound}]
Let $q>0$; there exist two positive constants $C,C_q$ such that, if $m=m_{4} + m_{2'} + m_{2}$:
\be
\sum_{\substack{\g\in\widetilde\TTT_{-1,m,q}  \\ n_{2',4}(\g) = n \\ m_{a}(\g) = m_{a}}}|S(\g)| \leq C^{m}C_{q}\|f\|_{1}^{q}n!\prod_{a}\Big[\max_{k\geq -1}|v_{k}^{(a)}|\Big]^{m_a}\;;\label{eq:nfat}
\ee
for $q=0$ the bound (\ref{eq:nfat}) has to be multiplied by $|\L|$.
\end{thm*}
We refer the reader to \cite{GN1, GN2, G}, and to appendix \ref{appD} (see remark 2 below), for a proof of this result. 
\vskip.1cm
{\bf Remarks.} 
\begin{enumerate}
\item The ``$n!$ bound'' (\ref{eq:nfat}) only applies to the planar theory; in the full theory $n$ is replaced by the number of endpoints of the tree. This proves the {\em ultraviolet stability} of the full $\varphi^{4}_4$ theory; see \cite{H, Z, DR, GN1, GN2, G, P}.
\item In references \cite{GN2, G} it was noticed that in the planar case the bound grows factorially in the number of frames; as we show in appendix \ref{appD}, it is possible to improve the bound by considering only the nontrivial frames labelled by $4$, $2'$. Roughly speaking, the factorial is ``produced'' by the sums appearing in the definitions of the frames; the frames labelled by $2,0$ do not contribute to the factorial because their sums can be controlled thanks to the exponential factor appearing in (\ref{1.4.1}) and figure \ref{fig02}, and if a frame is trivial the sum is missing.
\end{enumerate}
\vskip.1cm
{\bf Notations.} From now on we shall set
\be
\l := \l_{-1}\;,\qquad \a:= \a_{-1}\;,\qquad \m:= \g^{-2}\m_{-1}\;;\label{1.6a}
\ee
moreover, we define
\be
\ul{\l} := \{\l_{k}\}_{k\geq 1}\;,\qquad \ul{\a} := \{\a_{k}\}_{k\geq 1}\;,\qquad \ul\m := \{\m_{k}\}_{k\geq 1}\;.\label{1.6b}
\ee
Notice that the definition (\ref{1.6b}) does not involve the running coupling constants on scale zero; in fact, for purely technical reasons the running coupling constants on scale zero have to be treated separately from those on scales $>0$. In particular, we first determine the r.c.c. on scale $>0$ as functions of those on scale $0$, and then we express the r.c.c. on scale $0$ as functions of the renormalized ones. The motivation of this procedure is connected with the fact that the properties of the beta function (\ref{1.4a})--(\ref{1.4c}), that will play a key role in our analysis, are true only for scales $k>0$. It is also convenient to introduce
\be
\ul{\xi_{k}} := (\xi_{2',k},\xi_{2,k}) := (\a_k,\m_k)\;,\qquad \ul\xi := (\a,\m)\;,\qquad \ul{\ul \xi} := \{\ul{\xi_{k}}\}_{k\geq 1}\;.
\ee
Finally, we define the sets $\mathcal{B}_{\d}$, $\mathcal{C}_{\d}$, $\WW_{\d,\t}$ with $\d>0$, $\t\in\big(0,\frac{\pi}{2}\big]$ in the following way, see figure \ref{fig0}:
\bea
&&\mathcal{B}_{\delta} := \Big\{z\in\mathbb{C}:|z| < \delta\Big\}\;,\qquad \CC_{\d} := \Big\{z\in\mathbb{C}:\Re z^{-1}>
\delta^{-1}\Big\}\;,\nn\\
&&\WW_{\d,\t} := \Big\{ z\in\CCC: |z|<\d\;, |\arg z|< \pi - \t \Big\}\;.\label{1.12}
\eea
\section{Borel summability of $\varphi^{4}_4$ planar theory}\lb{sec3}
\renewcommand{\theequation}{\ref{sec3}.\arabic{equation}}
In this section we state our main result in a mathematically precise form, we recall what Borel summability is and we outline the ideas of the proof; the technical details are contained in section \ref{sec4} and in the appendices.
\begin{thm}['t Hooft -- Rivasseau]\label{th:1}
For any $\t\in\big(0,\frac{\pi}{2}\big]$ there exist $\bar\eta> 0$, $\bar\e>0$ such that the Schwinger functions $S^{T}(f;q) = \lim_{N\rightarrow+\infty} S^{T}_{(N)}(f;q)$ of the planar
$\varphi^{4}_{4}$ theory are analytic for $(\lambda,\alpha,\mu)\in\WW_{\bar\e,\t}\times\mathcal{B}_{\bar\eta}\times
\mathcal{B}_{\bar\eta}$, and Borel summable in $\l$ at the origin.
\end{thm}
\vskip.2cm
{\bf Remark.} Not surprisingly, $\bar\e\rightarrow 0$ if $\t\rightarrow 0$.
\vskip.2cm
Before discussing a sketch of the proof, let us briefly remind what Borel summability is (see \cite{Hdivser, S}). A formal power series $\sum_{n}a_{n} z^{n}$, $z\in\CCC$, is said to be Borel summable if the following properties are true:
\begin{itemize}
\item the {\em Borel transform} $B(t) := \sum_{n}\frac{a_n}{n!}t^{n}$ converges for every $t$ in some circle $\BB_{\d}$;
\item $B(t)$ admits an analytic continuation in a neighbourhood of the positive real axis;
\item the integral
\be
f(z) = \frac{1}{z}\int_{0}^{+\infty} e^{-\frac{t}{z}}B(t)\,dt \lb{1.13a}
\ee
is convergent for $z\in \CC_{\bar \d}$ for some $\bar \d>0$.
\end{itemize}
Notice that $f(z) \sim \sum_{n}a_{n}z^{n}$ for $z\rightarrow 0$. The function $f(z)$ is called the {\em Borel sum} of the formal power series, and if $f(z)$ exists it is {\em unique}. Therefore, Borel summability is nothing else than a one -- to -- one mapping between a certain space of functions and a certain space of power series: all the information on the function is enclosed in the list of its Taylor coefficients; for these reasons, Borel summability is, \cite{R3}, the perfect substitute for ordinary analyticity when a function is expanded on the boundary of its analyticity domain.

By Nevanlinna--Sokal theorem, \cite{S}, to establish whether $f(z)$ is the Borel sum of $\sum_{n}a_{n}z^{n}$ it is sufficient to check the following two properties:
\begin{itemize}
\item $f(z)$ is analytic in $\CC_{\d}$ for some $\d>0$;
\item for every $z\in\CC_{\d}$ and for all $M> 0$ the following estimate holds:
\be
\Big| f(z) - \sum_{n=0}^{M-1}a_{n}z^{n} \Big| \leq C^{M}M!|z|^{M}\;,\qquad C>0\;.\label{1.13}
\ee
\end{itemize}
\vskip.2cm
{\bf Sketch of the proof.} Our proof consists in a check of the two hypothesis of the Nevanlinna--Sokal theorem, and it goes as follows. First, we prove that for any fixed ultraviolet cutoff $N>0$ and any $\t\in\big(0,\frac{\pi}{2}\big]$ the running coupling constants are analytic for $(\l,\a,\m)\in \WW_{\bar\e,\t}\times\mathcal{B}_{\bar\eta}\times
\mathcal{B}_{\bar\eta}$; analyticity of the Schwinger functions $S^{T}_{(N)}(f;q)$ in the same domain is straightforward, since $S^{T}_{(N)}(f;q)$ is given by an absolutely convergent power series in the running coupling constants, see \cite{GN1, GN2, G}. Then, we prove that the limit $S^{T}(f;q)$ exists, and that it is reached uniformly in the analyticity domain; therefore, the limit is analytic in the same analyticity domain of $S^{T}_{(N)}(f;q)$. To conclude, we show that $S^{T}(f;q)$, as function of $\l$ in $\WW_{\bar\e,\t}$, verifies the bound (\ref{1.13}). These two properties imply Borel summability, since $\CC_{\bar\e}\subset \WW_{\bar\e,\t}$.
\vskip.1cm
{\it Analyticity.} To solve the flow equations (\ref{1.3c}) and determine the analyticity properties of the running coupling constants we use a fixed point argument. More precisely, we show that the equations (\ref{1.3c}) are solved by sequences parametrized by the renormalized coupling constants $(\l,\a,\m)$ which, for finite $N$, are the fixed points of some operators acting on suitable finite dimensional spaces; all the technical work is reduced to showing that in the considered spaces the operators are {\em contractions}. After this, the sequences of running coupling constants are determined through an exponentially convergent procedure. In particular, in the limit $N\rightarrow+\infty$, for $(\l,\a,\m)\in \WW_{\bar\e,\t}\times\BB_{\bar\eta}\times\BB_{\bar\eta}$, we find that the equations (\ref{1.3c}) admit a solution of the form, for some positive $C,c$:
\bea
&&\l_{k} = \frac{1}{\tilde\l^{-1} + \sum_{j=0}^{k}\tilde\b_k}\;,\quad |\a_{k} - \a| \leq c(|\l| + |\m|^2)\;,\nn\\&&\big|\m_{k} - \g^{-2k}\m \big| \leq c\big[\g^{-2k}|\m|^2 + (|\l| + |\ul\xi|) |\l_{k}|\big]\;,\label{1.6c}
\eea
where $\tilde\l = \l(1 + O(\m))$, $|\tilde\b_k - \b_k| \leq C(|\l|+|\ul\xi|)$, $\b_k := \b^{(4)}_{4,4}(k;k,k)>0$.

To begin, we rewrite the flow equation for $\l_k$ as, see (\ref{1.3c}) with $a=4$:
\bea
&&\l_{k} =: \l_{k+1}+\b_{4,k+1}\big(\ul\l,\ul{\ul \xi}\big)\;,\qquad k\geq 0\;,\label{1.7}\\
&&\l =: \l_{0} + f_{4,0}(\l_0,\m_0) + \b_{4,0}\big(\l_0,\ul{\l},\ul{\xi_0},\ul{\ul \xi}\big)\;,\label{1.7a}
\eea
where $f_{4,0}$ is linear in $\l_0$, and $\b_{4,h}$ is given by a sum of terms proportional to at least two among $\l_0,\ldots ,\l_N$. Then, we rewrite (\ref{1.3c}) for $a=2,2'$ as:
\bea
&&\a_{k} =: \alpha_0 - \sum_{j=1}^{k}\b_{2',j}\big(\ul{\l},\ul{\ul \xi}\big)\;,\qquad \m_{k} =: \g^{-2k}\m_0 - \sum_{j=1}^{k}\g^{2(j-k)}\b_{2,j}\big(\ul{\l},\ul{\ul \xi}\big)\;,\qquad k\geq 1\;,\qquad\label{1.8}\\
&&\a_0 =: \a - f_{2',0}(\l_0,\m_0) - \b_{2',0}\big( \l_0,\ul{\l},\ul{\xi_0},\ul{\ul \xi} \big)\;,\qquad \m_{0} =: \m - f_{2,0}(\m_0) - \b_{2,0}\big(\l_0,\ul{\l},\ul{\xi_0},\ul{\ul \xi} \big)\;,\qquad\label{1.8b}
\eea
where $f_{2',0}$ collect terms at most linear in $\l_0$, while $\b_{2',h}$, $\b_{2,h}$ are given by sums of terms proportional to at least two or one among $\l_0,\ldots , \l_N$, respectively. Setting $\b_{4,h}(\ul\l,\ul{\ul \xi}) =: \b_{h}\l_h^{2} + \bar\b_{4,h}\big(\ul{\l},\ul{\ul \xi}\big)$ where $\b_{h}>0$ and $\bar\b_{4,h}$ is of order $\geq 3$, equation (\ref{1.7}) can be rewritten as
\be
\l_{k}^{-1} = \l_{k+1}^{-1}-\b_{k+1}+R_{k+1}\big(\ul{\l},\ul{\ul \xi}\big)\Rightarrow \l_{k}^{-1} = \l_0^{-1} + \sum_{j=1}^{k}\b_{j} - \sum_{j=1}^{k}R_{j}\big(\ul{\l},\ul{\ul \xi}\big)\;,\lb{1.9}
\ee
where $R_{j}$ is given by a sum of terms proportional to one between $\a_j$, $\m_j$, $\l_j$, and it depends only on running coupling constants on scales $\geq j$, see appendix \ref{app:B}; the key remark is that, formally, equation (\ref{1.9}) can be seen as defining the {\em fixed point} of the map
\be
\left(\TT_{\l_{0},\ul{\ul \xi}}\,\ul{x}\right)_{k}=\frac{1}{\l_0^{-1}+\sum_{j=1}^{k}\b_{j}
-\sum_{j =
1}^{k}R_{j}\big(\underline{x},\ul{\ul\xi}\big)}\qquad\qquad k\geq 1\;,\label{1.10}
\ee 
where $\ul{x} = (x_1,...,x_N)$ with $x_i\in \CCC$ and $\underline{\alpha}$, $\underline{\mu}$ satisfy (\ref{1.8}), which again can be formally seen as the fixed point of the map
\begin{equation}
\left(\widetilde\TT_{\ul{\xi_0},\underline{\lambda}}\,\ul{\ul{y}}\right)_{k}=\left(\begin{array}{c} \alpha_0
- \sum_{j=1}^{k}\b_{2',j}\big(\underline{\lambda},\ul{\ul{y}}\big)\\
\gamma^{-2k}\mu_0 -
\sum_{j=1}^{k}\gamma^{2(j-k)}\b_{2,j}\big(\underline{\lambda},\ul{\ul{y}}\big)\end{array}\right)\qquad\qquad k\geq 1\;,\label{1.11}
\end{equation}
where $\ul{\ul{y}}=\left(\ul{y_1},\ul{y_2},\cdots,\ul{y_N}\right)$ and $\ul{y_k}=(y_{k,2'}\,,y_{k,2})$ with $y_{k,i}\in \CCC$.
Therefore, we can in principle determine the running coupling constants on scale $>0$ as functions of $(\l_{0},\,\a_{0},\,\m_{0})$ by solving the equations:
\be
\ul{\lambda} = \TT_{\l_{0},\ul{\ul{\xi}}}\,\ul{\l}\;,\qquad
\ul{\ul{\xi}} = \widetilde\TT_{\ul{\xi_0},\ul{\lambda}}\,\ul{\ul{\xi}}\;;\label{1.13b}
\ee
after this, the dependence of the running coupling constants on the renormalized ones can be deduced from equations (\ref{1.7a}), (\ref{1.8b}). 

To solve (\ref{1.13b}), in section \ref{sec:lemmi} and in appendices \ref{app:A}, \ref{app:B} we prove that if $\SS\in \CCC^{N}$ is the set of sequences ``close enough'' to the solution of the flow of $\l_k$ truncated to second order and if $\widetilde \SS\in \CCC^{2N}$ is a $2N$-dimensional ball centered in zero and of suitably small radius, then: (i) if $\ul x\in \SS$ and $|\a_0|$, $|\m_0|$ are small enough the map $\widetilde\TT_{\ul{\xi_0},\ul x}$ leaves $\widetilde\SS$ invariant and is a contraction therein; (ii) the fixed point $\ul{\ul y}(\ul x)$ of $\widetilde\TT_{\ul{\xi_0},\ul x}$ in $\widetilde\SS$ is H\"older continuous in $\ul{x}$ with exponent $0<\r<1$; (iii) given $\t \in (0,\pi/2]$, for all $\l_0\in \WW_{\e,\vartheta}$ with $\e$ small enough, the map $\TT_{\l_0,\ul{\ul y}(\cdot)}$ leaves $\SS$ invariant and is a contraction therein. To be specific, the distances $d$, $\tilde d$ that we shall adopt in $\SS$, $\widetilde\SS$ are defined as $d(\ul x,\ul x') := \max_k|x_k - x'_{k}|$, $\tilde d(\ul{\ul y},\ul{\ul y}') := \max_{k,i}|y_{k,i} - y'_{k,i}|$, respectively. 

Then, we can {\em construct} the sequences solving (\ref{1.13b}) in the following way: take
\be
\ul{\ul{\xi}}^{(0)} =
\left(\begin{array}{c}\underline{\alpha}^{(0)}\\\underline{\mu}^{(0)}\end{array}\right)\in\widetilde\SS,\qquad\underline{\lambda}^{(0)}\in\SS,\label{1.17}
\ee
and define, for $m\geq 0$,
\be
\ul{\ul{\xi}}^{(m+1)} := \lim_{n\rightarrow\infty}
\big(\widetilde\TT_{\ul{\xi_0},\ul{\lambda}^{(m)}}\big)^{n}\ul{\ul{\xi}}^{(m)}\;,\qquad \ul{\l}^{(m+1)}:= \TT_{\l_{0},\ul{\ul{\xi}}^{(m+1)}} \ul{\l}^{(m)}\;.\label{1.18}
\ee
Assume inductively that for all $0\leq m'\leq m$ the sequences $\ul{\ul{\xi}}^{(m')}$, $\ul{\l}^{(m')}$ belong respectively to $\widetilde\SS$, $\SS$, which is true for $m=0$. Property (i) above implies that $\ul{\ul{\xi}}^{(m+1)}$ belongs to $\widetilde\SS$, while property (iii) implies that $\ul{\l}^{(m+1)}$ belongs to $\SS$. Then, our procedure (\ref{1.18}) converges exponentially to a limit; in fact, for $m\geq 1$, for some $0<\r <1$, $C_{\r}>0$ and $0<\epsilon<1$:
\bea
&&\max_{k,i}\big|\xi^{(m+1)}_{k,i} - \xi^{(m)}_{k,i}\big|\leq C_{\r}\left(\max_{k}\big|\l_{k}^{(m)} - \l_{k}^{(m-1)}\big|\right)^{\r} \leq C_{\r}\epsilon^{(m-1)\r}\left(\max_{k}\big|\l_{k}^{(1)} - \l_{k}^{(0)}\big|\right)^{\r}\nn\\
&&\max_{k,i}\big|\l_{k}^{(m+1)} - \l_{k}^{(m)}\big| \leq \epsilon^{m}\max_{k}\big|\l_{k}^{(1)} - \l_{k}^{(0)}\big|\label{1.19}
\eea
where we used property (ii) to get the first inequality in the first line, and property (iii) for the remaining ones. Since $\ul{\l}^{(1)}$, $\ul{\l}^{(0)}$ are bounded, equations (\ref{1.19}) prove that the limits
\begin{equation}
\ul{\lambda}^{*} = \lim_{m\rightarrow\infty}\ul{\lambda}^{(m)},\qquad
\ul{\ul{\xi}}^{*} = \lim_{m\rightarrow\infty}\ul{\ul{\xi}}^{(m)}\label{1.20}
\end{equation}
exist in $\SS,\widetilde\SS$ respectively, and by construction
\begin{equation}
\underline{\lambda}^{*} = \TT_{\l_{0},\ul{\ul{\xi}}^{*}}\underline{\lambda}^{*},\qquad
\ul{\ul{\xi}}^{*} = \widetilde\TT_{\ul{\xi_0},\ul{\lambda}^{*}}\ul{\ul{\xi}}^{*}\;,\label{1.21}
\end{equation}
{\em i.e.} $\ul{\l}^{*}$, $\ul{\ul \xi}^{*}$ are the sequences of running coupling constants from scale $1$ to $N$ of the planar $\varphi^{4}_{4}$ theory, parametrized by $\l_0$, $\a_0$, $\m_0$. The proof of analyticity of the limits for $(\l_0,\a_0,\m_0)\in \WW_{\e,\t}\times \BB_{\eta}\times \BB_{\eta}$ with $\e$, $\eta$ small enough is straightforward; it is a consequence of the analyticity properties of the initial data and of the maps $\TT$, $\widetilde \TT$, and of the fact that convergence is uniform for $(\l_0, \a_0, \m_0)\in \WW_{\e,\t}\times\BB_{\eta}\times\BB_{\eta}$. After this, from equations (\ref{1.7a}), (\ref{1.8b}) we show that $\l_{0}$, $\a_{0}$, $\m_{0}$ are analytic for $(\l,\a,\m)\in \WW_{\e',\t'}\times \BB_{\eta'}\times \BB_{\eta'}$ with $\t'>\t$, $\e'<\e$, $\eta'<\eta$, and this concludes the proof of analyticity of the running coupling constants in the renormalized ones. 

Finally, to prove analyticity of the Schwinger functions we use that $S^{T}_{(N)}(f;q)$ is given by an absolutely convergent power series in the running coupling constants, see section \ref{sec2}, and we prove that the limit for  $N\rightarrow\infty$ exists and it is reached uniformly for $(\l,\a,\m)\in\WW_{\bar\e,\t'}\times\BB_{\bar\eta}\times\BB_{\bar\eta}$ with $\bar\e<\e'$, $\bar\eta<\eta'$.
\vskip.2cm
{\it Bound on the remainder.} In section \ref{sec4b} we show that relying on the tree representation of the beta function described in section \ref{sec2} it is possible to rewrite the $q$-point Schwinger function as:
\begin{equation}
S^{T}(f;q) = S^{T,(\leq n)}(f;q) + r^{(n)}(f;q)\;,\label{1.22}
\end{equation}
where $S^{T,(\leq n)}(f;q)$ is the Taylor expansion of $S^{T}(f;q)$ up to order $n$ in $\lambda=0$, and $r^{(n)}(f;q)$ is a quantity bounded by $(\const.)^{n+1}C_q\|f\|_1^{q}(n+1)!|\lambda|^{n+1}$ uniformly in the analyticity domain. The idea is to use the graphical representation of the beta function depicted in figure \ref{fig03} to ``extract'' in the tree expansion of the Schwinger function all the possible trees with less than $n+1$ thin endpoints corresponding to $\l$, as suggested by figure \ref{fig04}; the main difficulty in this procedure is to check that after having reproduced the Taylor series up to the order $n$ the ``unwanted'' trees, {\em i.e.} the ones showing more than $n$ endpoints of type $4$, have less than $n+1$ nontrivial frames labelled by $a=2',4$, see remark after figure \ref{fig04}. After having checked this, the desired bound is a straightforward consequence of the $n!$ bound (\ref{eq:nfat}).
\section{Proof of Theorem \ref{th:1}}\lb{sec4}
\renewcommand{\theequation}{\ref{sec4}.\arabic{equation}}

\subsection{Analyticity of the flow of the running coupling constants}\label{sec:lemmi}
In this section we present in a mathematically precise form the properties (i) -- (iii) mentioned in the previous section after equation (\ref{1.13b}), which, as we already discussed, are the key ingredients in the construction of the sequences of the running coupling constants on scale $\geq 1$ as functions of the ones on scale $0$. After this, we express the running coupling constants on scale $0$ in terms of the renormalized ones, and we prove the analyticity properties required for Borel summability. 

The spaces of sequences that we shall consider are the following ones:
\be
\SS_{\l_0,\d} := \Big\{\underline{x}\in \CCC^{N}:x_{k} = \frac{1}{\l_{0}^{-1} + \sum_{j=1}^{k}\b_{j} + t_k}\,,\, |t_k|\leq \sqrt \d k\Big\}\;,\quad \widetilde\SS_{\eta} := \left\{\ul{\ul{y}}\in \CCC^{2N}:|y_{k,i}|\leq\eta \right\} \label{2.1}
\ee
The following two lemmas imply respectively properties (i), (ii) and property (iii) stated in section \ref{sec3}.
\begin{lem}\label{lem:1}
For any $\t\in\big(0,\frac{\pi}{2}\big]$ there exist $\bar \e>0$, $\bar \eta>0$ such that if $(\l_0,\a_0,\m_0)\in \WW_{2\bar\e,\frac{\t}{2}}\times \BB_{2\bar\eta}\times \BB_{2\bar\eta}$ and $\underline{x}\in\SS_{\l_0,\bar\e + \bar\eta}$: 
\begin{enumerate}
\item $\widetilde\TT_{\ul{\xi_0},\underline{x}}$ is a map from $\widetilde\SS_{4\bar\eta}$ to
$\widetilde\SS_{4\bar\eta}$;
\item $\widetilde\TT_{\ul{\xi_0},\underline{x}}$ is a {\em contraction} in
$\widetilde\SS_{4\bar\eta}$, {\em i.e.} if $\ul{\ul{y}}\in\widetilde\SS_{4\bar\eta}$, $\ul{\ul y}' \in \widetilde\SS_{4\bar\eta}$
\be
\max_{k,i}\Big|\Big(\widetilde\TT_{\ul{\xi_0},\underline{x}}\,\ul{\ul{y}}\Big)_{k,i} - \Big(\widetilde\TT_{\ul{\xi_0},\underline{x}}\,\ul{\ul{y}}'\Big)_{k,i}\Big|\leq
\epsilon \max_{k,i}\big|y_{k,i} - y'_{k,i}\big|,\qquad 0<\epsilon<1;\label{2.3}
\ee
\item given two sequences $\underline{x}$, $\underline{x}'$ belonging to
$\SS_{\l_0,\bar\e + \bar\eta}$, the fixed points $\ul{\ul{y}}(\underline{x})$,
$\ul{\ul{y}}(\underline{x}')$ of the maps $\widetilde\TT_{\ul{\xi_0},\underline{x}}$,
$\widetilde\TT_{\ul{\xi_0},\underline{x}'}$ in $\widetilde\SS_{4\bar\eta}$ verify the following
inequalities:
\bea
\left|y_{k,i}(\underline{x})-y_{k,i}(\underline{x}')\right|&\leq&
C \big[\log(1+\bar\e k) +
1\big]\max_{k}|x_{k}-x'_{k}|\;,\label{eq:reg1}\\
\left|y_{k,i}(\underline{x})-y_{k,i}(\underline{x}')\right| &\leq& C_{\r}\Big(\max_{k}\big| x_{k} - x'_{k} \big|\Big)^{\r}\;.\lb{reg1b} 
\eea
for some positive $C$, $C_{\r}$ and $0<\r<1$.
\end{enumerate}
\end{lem}
\begin{lem}\label{lem:2}
For any $\t\in\big(0,\frac{\pi}{2}\big]$ there exist $\bar\e>0$, $\bar{\eta}>0$ such that if $(\l_0,\a_0,\m_0)\in \WW_{2\bar\e,\frac{\t}{2}}\times \BB_{2\bar\eta}\times \BB_{2\bar\eta}$ the fixed point $\ul{\ul y}(\ul x)$ of $\widetilde\TT_{\ul{\xi_0},\ul x}$ in $\widetilde\SS_{4\bar\eta}$ for $\ul{x}\in\SS_{\l_0,\bar\e + \bar\eta}$ exists and:
\begin{enumerate}
\item $\TT_{\l_{0},\ul{\ul{y}}(\cdot)}$ is a map from $\SS_{\l_0,\bar\e + \bar\eta}$ to
$\SS_{\l_0,\bar\e + \bar\eta}$;
\item $\TT_{\l_{0},\ul{\ul{y}}(\cdot)}$ is a {\em contraction} in
$\SS_{\l_0,\bar\e + \bar\eta}$, {\em i.e.} if $\ul{x}\in\SS_{\l_0,\bar\e + \bar\eta}$, $\ul{x}'\in\SS_{\l_0,\bar\e + \bar\eta}$,
\be
\max_{k}\Big|\Big(\TT_{\l_{0},\ul{\ul{y}}(\ul x)}\,\ul{x}\Big)_{k} - \Big(\TT_{\l_{0},\ul{\ul{y}}(\ul x')}\,\ul{x}'\Big)_{k}\Big|\leq
\epsilon \max_{k}\big| x_{k} - x'_{k}\big|\;,\qquad 0<\epsilon<1.\label{2.6}
\ee
\end{enumerate}
\end{lem}
We refer the reader to appendices \ref{app:A}, \ref{app:B} for the proofs of these lemmas. As explained in section \ref{sec3}, this two results allow to construct the sequences
of the running coupling constants as functions of those on scale $0$, and to
determine their analyticity properties. We take
\be
\ul{\ul{\xi}}^{(0)} =
\left(\begin{array}{c}\underline{\alpha}^{(0)}\\\underline{\mu}^{(0)}\end{array}\right)\in\widetilde\SS_{4\bar{\eta}},\qquad\underline{\lambda}^{(0)}\in\SS_{\l_0,\bar\e + \bar\eta}\label{2.7}
\ee
analytic for $(\lambda_0,\alpha_0,\mu_0)\in \WW_{2\bar\e,\frac{\t}{2}}\times
\BB_{2\bar\eta}\times
\BB_{2\bar\eta}$; to be
concrete, we can choose
\begin{equation}
\alpha^{(0)}_{k} = \mu^{(0)}_{k} = \bar\eta,\qquad
\lambda^{(0)}_{k} = \frac{1}{\l_0^{-1} + \sum_{j=1}^{k}\b_{j}}\;,\qquad \l_{0}\in \WW_{2\bar\e,\frac{\t}{2}}\;.\label{2.8}
\end{equation}
Then, we can construct the sequences of running coupling constants by proceeding as explained after (\ref{1.17}); analyticity for $(\l_0,\a_0,\m_0)\in \WW_{2\bar\e,\frac{\t}{2}}\times \BB_{2\bar\eta}\times\BB_{2\bar\eta}$ is a straightforward consequence of the analyticity properties of the maps and of the initial data, and of the fact that convergence is uniform for $(\l_0,\a_0,\m_0)\in \WW_{2\bar\e,\frac{\t}{2}}\times \BB_{2\bar\eta}\times\BB_{2\bar\eta}$.

Now we turn to the flow equations (\ref{1.7a}), (\ref{1.8b}) for the running coupling constants on scale $0$. Notice that these equations are {\em different} from the ones corresponding to higher scales, because of the presence of the functions $f_{a,0}$. The main consequence of this fact is that choosing $\l$ inside $\CC_{\e}$ {\em does not} imply that $\l_{0} \in \CC_{\e'}$ for some $\e'$; this is the reason why we considered $\l_0\in \WW_{\e,\vartheta}$ so far. The strategy that we shall adopt is very similar, but technically much simpler, to the one we followed for the scales $1,...,N$, see appendix \ref{appscale0} for details: first, we determine with a fixed point argument $\a_0$, $\m_0$ as analytic functions of $\l_0$, $\a$, $\m$ in $\WW_{2\e,\frac{\t}{2}}\times \BB_{\eta}\times\BB_{\eta}$ for $\e$, $\eta$ small enough; then, we plug $\a_0$, $\m_0$ into the equation (\ref{1.7a}) for $\l_0$, and we solve it using again a fixed point argument; finally, we show that the solution has the required analyticity properties in $\l$, $\a$, $\m$. In particular, it follows that $\l_0^{-1} \simeq \l^{-1}(1 + O(\m)) + \b_{0}$, up to corrections bounded by $(\const.)(|\l| + |\ul{\xi}|)$.
\vskip.2cm
{\it Asymptotic behaviour of the running coupling constants.} So far, our construction allowed us to conclude that, if $(\l,\a,\m)\in \WW_{\e,\t}\times\BB_{\eta}\times\BB_{\eta}$ with $\e$, $\eta$ small enough:
\be
\l_{k} = \frac{1}{\l^{-1}(1 + O(\m)) + \sum_{j=0}^{k}\b_k + t'_k}\;,\qquad |\a_{k}|\leq \eta\;,\qquad |\m_{k}|\leq \eta\;, \lb{2.9a}
\ee
with $|t'_k|\leq (k+1)\sqrt{\e+\eta}$; however, these results can be improved to get (\ref{1.6c}). In fact, the flows of $\a_{k}$, $\m_{k}$ are given by, for $k\geq 1$:
\be
\a_{k} = \a_0 - \sum_{j=1}^{k}\b_{2',j}\big(\ul{\l},\ul{\ul \xi}\big)\;,\qquad \m_{k} = \g^{-2k}\m_0 - \sum_{j=1}^{k}\g^{2(j-k)}\b_{2,j}\big(\ul{\l},\ul{\ul \xi}\big)\;,\lb{2.9b}
\ee
where:
\bea
\big| \b_{2',j}\big(\ul{\l},\ul{\ul \xi}\big) \big| \leq c'|\l_{j}|^{2}\;,\qquad \big| \b_{2,j}\big(\ul{\l},\ul{\ul \xi}\big) \big| \leq c' \big(|\l| + |\ul\xi|\big) |\l_{j}|\;.\lb{2.9c}
\eea
Therefore it follows that, using the expression for $\l_k$ in (\ref{2.9a}), for some $c>0$:
\be
|\a_{k} - \a| \leq c\big(|\l| + |\m|^2\big)\;,\qquad \big|\m_{k} - \g^{-2k}\m\big| \leq c\big[\g^{-2k}|\m|^2 + (|\l| + |\ul\xi|) |\l_{k}|\big]\;,\lb{2.9d}
\ee
which give the last two of (\ref{1.6c}). To prove the first of (\ref{1.6c}), simply use (\ref{2.9d}) and the first of (\ref{2.9a}) to replace the running coupling constants appearing in $R_j$, see (\ref{1.10}) and (\ref{B0}).
\vskip.2cm
{\it Analyticity of the Schwinger functions.} As we have discussed in section \ref{sec2}, the Schwinger functions $S^{T}_{(N)}(f;q)$ are given by absolutely convergent power series in the running coupling constants on scales $\leq N$; therefore, taking $\bar\e$, $\bar\eta$ smaller than the radius of convergence of the series, $S^{T}_{(N)}(f;q)$ is analytic for $(\l,\a,\m)\in \WW_{\bar\e,\t}\times
\mathcal{B}_{\bar\eta}\times\mathcal{B}_{\bar\eta}$. To prove analyticity in the limit $N\rightarrow+\infty$ we show that the sequence $\big\{S^{T}_{(N)}(f;q)\big\}_{N\geq 1}$ is uniformly Cauchy in the analyticity domain. In fact, consider two positive integers $N,N'$ such that $N'>N$; then,
\be
S^{T}_{(N')}(f;q) - S^{T}_{(N)}(f;q) := \d S_{1,(N,N')}^{T}(f;q) + \d S_{2,(N,N')}^{T}(f;q)\;,\lb{2.8a}
\ee 
where: $\d S_{1,(N,N')}^{T}(f;q)$ is given by a sum of trees with at least one endpoint on scale $k \leq N$ corresponding to the difference of running coupling constants $v_{k}^{(a),N'} - v_{k}^{(a),N}$ of theories with cut--offs on scales $N'$, $N$, and $\d S_{2,(N,N')}^{T}(f;q)$ is given by a sum of GN trees having root scale $-1$ and at least one endpoint on scale $\geq N+1$. The first term can be bounded using the results of appendix \ref{appuv} as:
\be
\left|\d S_{1,(N,N')}^{T}(f;q)\right|\leq (\const.)N^{-1}\;,\lb{2.8ab}
\ee
while the second can be estimated using the short memory property of the GN trees (see discussion after (\ref{1.3.7})) as, for some $\r>0$,
\be
\left| \d S_{2,(N,N')}^{T}(f;q)\ \right| \leq (\const.)\g^{-\r N}\;,\qquad \g>1\;;\lb{2.8ac}
\ee
all the bounds are uniform in $(\l,\a,\m) \in \WW_{\bar\e,\t}\times
\mathcal{B}_{\bar\eta}\times\mathcal{B}_{\bar\eta}$. Therefore the limit exists, and it is analytic in $\WW_{\bar\e,\t}\times
\mathcal{B}_{\bar\eta}\times\mathcal{B}_{\bar\eta}$. 
 
\subsection{Bounds on the Taylor remainder of the Schwinger functions}\label{sec4b}

In this section we show that for all $n>0$, $(\lambda,\alpha,\mu)\in
\WW_{\bar\e,\t}\times \mathcal{B}_{\bar\eta}\times
\mathcal{B}_{\bar\eta}$, the $q$ -- points Schwinger function $S^{T}(f;q)$ verifies
\be
S^{T}(f;q) = S^{T,(\leq n)}(f;q) + r_{n}(f;q)\lb{4.1}
\ee
where $S^{T,(\leq n)}(f;q)$ is the Taylor expansion of $S^{T}(f;q)$ up to the order $n$ in $\l=0$ and $r_{n}(f;q)$ is a remainder bounded by $C^{n+1} (n+1)!|\l|^{n+1}$ for some $C>0$; result (\ref{4.1}) concludes the proof of Borel summability of the Schwinger functions of the planar theory. 

\begin{figure}[htbp]
\centering
\includegraphics[width=0.8\textwidth]{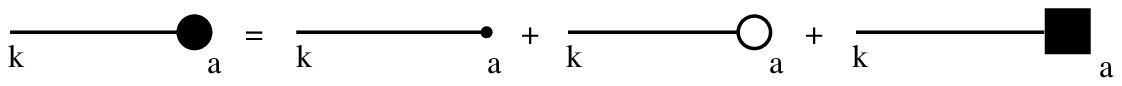}
\caption{Fat endpoints are equal to thin plus empty plus square endpoints.} \label{fig05}
\end{figure}

One can try to prove decomposition (\ref{4.1}) by iterating the graphical definition of the running coupling constants, see discussion after (\ref{1.4.1}) and in particular figure \ref{fig04} to get an idea of the graphical meaning of the iteration, to ``extract'' all the possible trees with only thin endpoints and at most $n$ of them labelled by $4$; to conclude the proof one has to check at the end that the sum of the values of the trees not belonging to this category is bounded by $C^{n+1}(n+1)!|\l|^{n+1}$. For simplicity, in the following we shall call ``$a$-endpoint'' an endpoint labelled by $a$, and ``$a$-frame'' a frame labelled by $a$; $a$-frames with $a$ equal to $2'$ or $4$ will be called ``$(2',4)$-frames''.
\vskip.2cm
{\em Empty and square endpoints.} We can rewrite (\ref{1.7}), (\ref{1.8b}) in the more compact form:
\be
v_{k}^{(a)} = \g^{-2\d_{a,2}(k+1)}v_{-1}^{(a)} - \g^{-2\d_{a,2}k}f_{a,0}(\l_0,\m_0) - \sum_{j=0}^{k}\g^{2(j-k)\d_{a,2}}\b_{a,j}\big(\l_0,\ul\l,\ul{\xi_0},\ul{\ul \xi}\big)\;.\label{4.1.0}
\ee
We graphically represent $- \g^{-2\d_{a,2}k}f_{a,0}$ as an {\em empty} $a$--endpoint and $- \sum_{j=0}^{k}\g^{2(j-k)\d_{a,2}}\b_{a,j}$ as a {\em square} $a$-endpoint; therefore, in general the fat $a$-endpoint can be written as the sum of thin, empty and square $a$-endpoints; see figure \ref{fig05}. In turn, the empty and the square endpoints can be represented as sums of framed trees with root scale $k$, see section \ref{sec2}; it is important to notice that the frames appearing in the tree representation of $- \g^{-2\d_{a,2}k}f_{a,0}$ are {\em trivial}, see remark after figure \ref{fig03}.

We define the {\em order} and the {\em $4$-order} of fat, thin, empty, and square endpoints as the order of their values in all the renormalized coupling constants and in $\l$ only, respectively. Therefore:
\begin{itemize}
\item Thin and fat endpoints have order $1$; empty endpoints have order $2$; square $a$-endpoints have order $1$ or $2$ depending on whether $a=2',4$ or $a=2$.
\item Thin, fat and empty $a$-endpoints have $4$-order $0$ or $1$ depending on whether $a=2',2$ or $a=4$; square endpoints have $4$-order $1$.
\end{itemize}
Notice that the reason why we set to $1$ the order and the $4$-order of the square $a$-endpoints with $a=2',4$, which are given by sums of trees with {\em two} $4$-endpoints, is that we have to exploit asymptotic freedom to control the sum in (\ref{4.1.0}); the result can be bounded uniformly in $k$ by $|\l|$ but not by $|\l|^2$.
\vskip.1cm
{\bf Notations.} We shall use the following notations:
\begin{itemize}
\item $n_{2',4}(\g)$ is the number of nontrivial $(2',4)$-frames appearing in a tree $\g$;
\item $n_{sq}^{(a)}(\g)$ is the number of square $a$-endpoints appearing in a tree $\g$, and $n_{sq}(\g) := n^{(4)}_{sq}(\g) + n^{(2')}_{sq}(\g) +n^{(2)}_{sq}(\g)$;
\item the order $O(\g)$ and the $4$-order $O_4(\g)$ of a tree $\g$ are respectively equal to the sums of the orders, $4$-orders of the endpoints of $\g$;
\item the ``expansion'' of square and empty endpoints consists in replacing them with their tree expansions.
\end{itemize}

{\it Proof of (\ref{1.22}).} We will proceed by induction. Assume that, at the step $r$ of the induction, for every $n>0, M>0$ with $M\geq n$ the Schwinger function $S^{T}(f;q)$ can be written as
\be
S^{T}(f;q) = \mathcal{F}_{n, M}^{(r)}+\mathcal{R}_{n,M}^{(r),1} + \RR_{n,M}^{(r),2}\;,
\ee 
where both $\FF_{n,M}^{(r)}$, $\RR_{n,M}^{(r),i}$ can be represented as sums over distinct trees such that $n_{2',4}(\g)\leq n$; moreover, we assume that:
\begin{itemize}
\item the trees $\g$ contributing to $\FF^{(r)}_{n,M}$ are such that $O_4(\g)\leq n$, $O(\g)\leq M$ and show fat and thin endpoints;
\item the trees $\g$ contributing to $\RR_{n,M}^{(r),i}$ are such that $O_4(\g)>n$ or $O(\g)>M$, depending on whether $i=1,2$, and may have empty and square endpoints.
\end{itemize}
These assumptions are trivially true at the beginning of the induction, see section \ref{sec2}. As a consequence of result (\ref{eq:nfat}), and since the number of topologically distinct trees with $m$ endpoints is estimated by $(\const.)^{m}$, $\mathcal{R}_{n,M}^{(r),1}$, $\RR_{n,M}^{(r),2}$ are bounded respectively by $C^{n}C_q\|f\|_{1}^{q}n!|\l|^{n+1}$,  $C^{M}C_q\|f\|_1^{q}n!\d^{M+1}$ for some positive $C$ and $\d := \max_{k}\{|\l_k|,|\a_k|,|\m_k|\}$. 
Now do the following.
\vskip.3cm
{\bf 1.} Substitute every fat $2$-endpoint appearing in $\mathcal{F}_{n, M}^{(r)}$ with the sum of a thin plus an empty plus a square $2$-endpoints: in this way the fat
$2$-endpoint disappear, generating new trees such that $n_{2',4}(\g)\leq n$ that we organize by writing
\be
\mathcal{F}_{n,M}^{(r)} = \AA_{1}^{(r)} +
\AA_{2}^{(r),1} + \AA_{2}^{(r),2}\;,
\ee
where
\begin{eqnarray}
\AA_{1}^{(r)} &:=& \mbox{``sum of trees $\gamma$ such that $O_{4}(\g) \leq n$ and $O(\g)\leq M$'',}\nonumber\\
\AA_{2}^{(r),1} &:=& \mbox{``sum of trees $\gamma$ such that $O_4(\g)>n$'',}\nonumber\\
\AA_{2}^{(r),2} &:=& \mbox{``sum of trees $\gamma$ such that $O_4(\g)\leq n$ and $O(\g)>M$''.}\nn
\end{eqnarray}
\vskip.2cm
{\bf 2.} Substitute every fat $2'$-endpoint appearing in $\AA_{1}^{(r)}$ with the sum of a thin plus an empty plus a square $2'$-endpoint: in this way the
fat $2'$-endpoints disappear, generating new trees such that $n_{2',4}(\g)\leq n$ that we organize by writing
\be
\AA_{1}^{(r)} = \AA_{3}^{(r)} +
\AA_{4}^{(r),1} + \AA_{4}^{(r),2}\;,
\ee
where,
\begin{eqnarray}
\AA_{3}^{(r)} &:=& \mbox{``sum of trees $\gamma$ s.t. $n_{2',4}(\gamma)+n_{sq}^{(2')}(\gamma)\leq
n-1 + \d_{n_{2',4},0}$ and $O(\g)\leq M$'',}\nonumber\\
\AA_{4}^{(r),1} &:=& \mbox{``sum of trees $\gamma$ s.t. $n_{2',4}(\gamma)+n_{sq}^{(2')}(\gamma) >
n-1 + \d_{n_{2',4},0}$'',}\nonumber\\
\AA_{4}^{(r),2} &:=&  \mbox{``sum of trees $\gamma$ s.t. $n_{2',4}(\gamma)+n_{sq}^{(2')}(\gamma) \leq
n-1 + \d_{n_{2',4},0}$ and $O(\g) > M$''.}\nn
\end{eqnarray} 
Notice that the trees appearing in $\AA_{4}^{(r),1}$ are such that $O_4(\g)>n$; in fact, for these trees,
\be
O_4(\g) \geq n_{2',4}(\g) + 1 + n_{sq}^{(2')}(\g) - \d_{n_{2',4},0} > n\;,\label{4.9b}
\ee
where we used that each nontrivial $2'$-frame contains trees of $4$-order $\geq 2$, that the square $2'$-endpoints are of $4$-order strictly bigger than their corresponding thin and empty endpoints, and the definition of $\AA_4^{(r),1}$. 
\vskip.2cm
{\bf 3.} Expand each square $a$-endpoint with $a=2',2$ appearing in $\AA_3^{(r)}$, and write
\begin{equation}
\AA_{3}^{(r)} = \AA_5^{(r)} + \AA_{6}^{(r),1} + \AA_{6}^{(r),2},
\end{equation} 
where
\bea
\AA_5^{(r)} &:=& \mbox{``sum of the trees $\g$ s.t. $O_4(\g)\leq n$ and $O(\g)\leq M$''}\;,\nn\\
\AA_6^{(r),1} &:=& \mbox{``sum of the trees $\g$ s.t. $O_4(\g)>n$''}\;,\nn\\
\AA_{6}^{(r),2} &:=& \mbox{``sum of the trees $\g$ s.t. $O_4(\g)\leq n$ and $O(\g)>M$''}.\nn
\eea
Notice that the trees generated at this step are such that $n_{2',4}(\g)\leq n$; in fact, for a generic tree $\g'$ generated by $\g\in \AA_{3}^{(r)}$ it follows that $n_{2',4}(\g') = n_{2',4}(\g) + n_{sq}^{(2')}(\g)\leq n$, where the last inequality holds by definition of $\AA_3^{(r)}$.
\vskip.2cm
{\bf 4.} Substitute every fat $4$-endpoint appearing in $\AA_5^{(r)}$ with the sum of a thin plus an empty plus a square $4$-endpoint: in this way the fat
$4$-endpoints disappear, generating new trees such that $n_{2',4}(\g)\leq n$ that we organize by writing
\begin{equation}
\AA_5^{(r)} = \AA_{7}^{(r)} + \AA_{8}^{(r),1} + \AA_{8}^{(r),2},
\end{equation}
where,
\begin{eqnarray}
\AA_{7}^{(r)} &:=& \mbox{``sum of trees $\gamma$ s.t. $n_{2',4}(\gamma)+n_{sq}^{(4)}(\gamma)\leq
n-1 + \d_{n_{2',4},0}$ and $O(\g)\leq M$'',}\nonumber\\
\AA_{8}^{(r),1} &:=& \mbox{``sum of trees $\gamma$ s.t. $n_{2',4}(\gamma)+n_{sq}^{(4)}(\gamma) >
n-1 + \d_{n_{2',4},0}$'',}\nonumber\\
\AA_{8}^{(r),2} &:=& \mbox{``sum of trees $\gamma$ s.t. $n_{2',4}(\gamma)+n_{sq}^{(4)}(\gamma) \leq
n-1 + \d_{n_{2',4},0}$ and $O(\g)>M$''.}\nonumber
\end{eqnarray} 
Now, we show $\AA_{8}^{(r),1}$ can be rewritten as a sum of trees such that $O_4(\g)>n$ and $n_{2',4}(\g)\leq n$. Notice that since the $4$-order of the $4$-square endpoint is equal to the $4$-order of its corresponding fat, thin and empty endpoints, we cannot use a bound like the one in (\ref{4.9b}). To ``rise the $4$-order'' of a tree $\g$ up to $n+1$ we have to expand a suitable number $\tilde n_{sq}^{(4)}(\g)\leq n_{sq}^{(4)}(\g)$ of square $4$-endpoints (which are given by sums of trees of $4$-order $\geq 2$). If $n_{2',4}(\g) = 0$ we choose $\tilde n_{sq}^{(4)}(\g) = 0$,  because in this case by definition of $\SS_{4}^{(r)}$ the $4$-order of $\g$ is already $>n$; if $n_{2',4}(\g)>0$ we choose
\be
\tilde n_{sq}^{(4)}(\g) := n - n_{2',4}(\g)\;,
\ee
with this choice it follows that ($n_{2',4}(\g)$ refers to the tree $\g$ before this last expansion),
\be
O_4(\g) \geq n_{2',4}(\g) + 1 + \tilde n_{sq}^{(4)}(\g) = n+1\;.
\ee
Finally, a generic tree $\g'$ produced by this last expansion verifies
\be
n_{2',4}(\g') =  n_{2',4}(\g) + \tilde n_{sq}^{(4)}(\g) = n\;.
\ee 
\vskip.2cm
{\bf 5.} Expand each square $4$-endpoint appearing in $\AA_{7}^{(r)}$, and write
\begin{equation}
\AA_{7}^{(r)} = \AA_9^{(r)} + \AA_{10}^{(r),1} + \AA_{10}^{(r),2}\;,
\end{equation} 
where
\bea
\AA_9^{(r)} &:=& \mbox{``sum of the trees $\g$ s.t. $O_4(\g)\leq n$ and $O(\g)\leq M$,''}\nn\\
\AA_{10}^{(r),1} &:=& \mbox{``sum of the trees $\g$ s.t. $O_4(\g)>n$,''}\nn\\
\AA_{10}^{(r),2} &:=& \mbox{``sum of the trees $\g$ s.t. $O_4(\g)\leq n$ and $O(\g)>M$.''}\nn\\
\eea
Notice that the trees generated at this step are such that $n_{2',4}(\g)\leq n$; in fact, for a generic tree $\g'$ generated by $\g\in \AA_{7}^{(r)}$ it follows that $n_{2',4}(\g') = n_{2',4}(\g) + n^{(4)}_{sq}(\g)\leq n$, where the last inequality holds by definition of $\AA_{7}^{(r)}$.
\vskip.2cm
{\bf 6.} Expand each empty $a$-endpoint appearing in $\AA_{9}^{(r)}$, and write
\bea
\AA_9^{(r)} = \mathcal{F}^{(r+1)}_{n,M} + \AA_{12}^{(r),1} + \AA_{12}^{(r),2}
\eea
where
\bea
\mathcal{F}^{(r+1)}_{n,M} &:=& \mbox{``sum of the trees $\g$ s.t. $O_4(\g)\leq n$ and $O(\g)\leq M$'',}\nn\\
\AA_{12}^{(r),1} &:=&  \mbox{``sum of the trees $\g$ s.t. $O_4(\g)>n$,''}\nn\\
\AA_{12}^{(r),2} &:=&  \mbox{``sum of the trees $\g$ s.t. $O_4(\g)\leq n$ and $O(\g)>M$.''}\nn\\
\eea
\vskip.2cm
{\bf 7.} We are now able to express the generic Schwinger function $S^{T}(f;q)$  as
\bea
S^{T}(f;q) &=& \mathcal{F}_{n,M}^{(r+1)} + \mathcal{R}_{n,M}^{(r),1} + \RR_{n,M}^{(r),2} + \sum_{j=1,2}\sum_{i=1}^{6}\AA_{2i}^{(r),j}\nn\\
                 &=:& \mathcal{F}_{n,M}^{(r+1)} + \mathcal{R}_{n,M}^{(r+1),1} + \mathcal{R}_{n,M}^{(r+1),2},\label{eq:2step1}
\eea
where, by construction, all the trees are such that $n_{2',4}(\g)\leq n$, the remainder $\mathcal{R}_{n,M}^{(r+1),1}$ contains distinct trees such that $O_4(\g)>n$, 
while $\RR_{n,M}^{(r+1),2}$ is given by a sum of distinct trees such that $O(\g)>M$. 
If $\mathcal{F}_{n,M}^{(r+1)}$ still contains trees with fat endpoints repeat
the process starting from step {\bf 1.}, otherwise we have finished: calling $r^{*}$ the
final step (which is finite, see remark (1) below), {\it i.e.} the integer such that $\mathcal{F}_{n,M}^{(r^{*})}$
contains trees with only thin endpoints, the $n!$ bound (\ref{eq:nfat}) implies that, if $\d = \max_{h}\{|\l_{h}|, |\a_{h}|, |\m_h|\}$:
\be
|\RR_{n,M}^{(r^{*}),1}| \leq C^{n}C_q\|f\|_{1}^{q}n!|\l|^{n+1}\;,\qquad |\RR_{n,M}^{(r^{*}),2}| \leq C^{M}C_q\|f\|_{1}^{q}n!\d^{M+1}\;.
\ee
Moreover, $\FF_{n,M}^{(r^{*})}$ differs from $\FF_{n,+\infty}^{(r^{*})}$, the Taylor expansion in $\l$ to the order $n$, by a quantity bounded by $C^{M}C_q\|f\|_{1}^{q}n!\d^{M+1}$; therefore, for each $\l$ in the analyticity domain and for each $n\geq 0$ there exists a finite integer $M(\l,n)\geq n$ such that for all $M\geq M(\l,n)$ it follows that:
\be
\big|S^{T}(f;q) - \FF_{n,+\infty}^{(r^{*})}\big|\leq 4C^{n}C_q\|f\|_{1}^{q}n!|\l|^{n+1}\;,
\ee
and this bound concludes the proof of Borel summability of the $\varphi^{4}_4$ planar theory.
\vskip.3cm
{\bf Remark}. The iteration ends in less than $M+1$ steps (where each step is formed by the seven
substeps described above); this means that no trees with fat
endpoints are present in $\mathcal{F}^{(M+1)}_{n,M}$. We can prove this fact with a simple induction. At the
step $r=0$ the trees with fat endpoints are of order $\geq 0$; assume inductively
that at the $r$-th step the trees belonging to $\mathcal{F}_{n,M}^{(r)}$ with
at least one fat endpoint are of order $\geq r$; if this is true, by repeating the
six substeps described above we find that the new trees with at least one fat 
endpoint appearing in $\mathcal{F}^{(r+1)}_{n,M}$ must be of order $\geq
r+1$, since at the $r$-th step the fat endpoints are replaced by thin plus empty plus square endpoints, and the empty endpoints are of order $2$ while
the squares are given by sums of trees of order $\geq 2$. 
Hence, after {\em at most} $r^{*} = M+1$ iterations no more trees showing fat endpoints will be present in $\mathcal{F}^{(r^{*})}_{n,M}$.

%To prove that we need {\em exactly} $M$ iterations to conclude the procedure we have
%to find at the $r$-th step a tree of order $r$ showing one or more fat endpoints.
%A tree with such features is the ``comb tree'' depicted in figure \ref{fig:3a}; this
%tree is generated by the tree at the l.h.s. of figure \ref{fig:3a}, which contributes to $S^{T}(f;2)$, and at each
%iteration grows in the ``slowest way'', {\it i.e.} by raising its order of only one.
%The number of iterations needed to grow a comb of order $n$ with only thin endpoints
%is equal to $M-1$.
%
%\begin{figure}[htbp]
%\centering
%\includegraphics[width=0.8\textwidth]{comb_S.eps}
%\caption{A ``comb tree'' of order $r$, produced at the $r$-th step of the
%iteration; this tree contributes to $S^{T}(f;2)$. All the frames labelled by $4$, except the first which is labelled by $2$; all the endpoints are of type $4$, except the first and the second from left which are labelled by $2$, and only the two rightmost endpoints are fat.} \label{fig:3a}
%\end{figure} 
%

\section{Conclusions}\label{sec5}
\renewcommand{\theequation}{\ref{sec5}.\arabic{equation}}

In this paper we discussed the issue of Borel summability in the framework of multiscale analysis and renormalization group, by providing a proof of Borel summability for the $\varphi^{4}_4$ planar theory using the techniques of \cite{G}.
This result is not new, since it has been proven independently by 't Hooft and Rivasseau, \cite{tH, tH2, R, R2}. The proof given by 't Hooft is based on renormalization group methods, and it does not rely on Nevanlinna -- Sokal theorem; we have not been able to fully reproduce 't Hooft argument in our rigorous framework. The proof given by Rivasseau, instead, consists in a check of the two hypothesis of Nevanlinna -- Sokal theorem; however, his methods are quite different from the ones that we use, since in his approach the beta function was not introduced. Moreover, in his work a particular choice of the wave function renormalization and of the renormalized mass was made. 

One of the motivations of our work is that very few proofs of Borel summability of interacting field theories based on renormalization group methods are present in the literature, \cite{FMRS, FMRS2}. Moreover, our framework has already been proved effective in the analysis of various models of condensed matter and field theory; therefore, we consider our work as a first step towards the analysis of more interesting models. For instance, we think that the ideas of this paper can be applied to the one dimensional Hubbard model, which has been rigorously constructed through renormalization group methods in \cite{M}, but where a proof of Borel summability has not been given yet. In fact, due to the anticommutativity of the fermionic fields the factorial growth of the Feynman graphs can be controlled using the so called {\em Gram bounds}; moreover, one sector of the theory is asymptotically free, while to control the flow of the other running coupling constants one has to exploit the vanishing of the beta function.

Regarding our work, the first part of this paper consists essentially in a rigorous study of the beta function of an asymptotically free field theory. In particular, we have shown that the theory is analytic for values of the renormalized coupling constant $\l$ belonging to a ``Watson domain'', see \cite{Watson} and definition (\ref{1.12}), and for values of the wave function renormalization and of the renormalized mass around $1$. In the second part of our work, to prove Borel summability we have shown that it is possible to ``undo'' the resummation that allowed us to write the Schwinger functions as a convergent power series in the running coupling constants, in such a way that the difference between the generic Schwinger function and its Taylor expansion to the order $n$ in $\l$ is bounded by $C^{n}n!|\l|^{n+1}$ for some positive $C$; thanks to Nevanlinna--Sokal theorem, see \cite{S}, this last fact along with the above mentioned analyticity properties implies Borel summability.
\vskip.2cm
\section{Acknowledgements}

It is a pleasure to thank Prof. G. Gallavotti for having introduced us to the theory of renormalization, for having proposed the problem and for many very useful discussions, from which all the ideas of this paper emerged. We are also grateful to Dr. A. Giuliani, for constant encouragement and constructive criticism.

%\begin{appendix}
%\include{appl}

\appendix

\section{Proof of lemma \ref{lem:1}}\label{app:A}
\renewcommand{\theequation}{\ref{app:A}.\arabic{equation}}

In this appendix we present the proof of lemma \ref{lem:1}. Recall that $\bar r$ is the number of running coupling constants of type $4$ appearing at a given order $r$ of the perturbative series defining the beta function, see (\ref{1.4}); moreover, we define $\tilde r := r - \bar r$. We remind also that with the notation $\ul{\ul{y}}(\underline{x})$ we denote the fixed point of the map $\widetilde\TT_{\ul{\xi_0},\ul{x}}$ in $\widetilde\SS_{2\eta}$.

All the estimates that we shall derive here and in the next appendix are consequences of the fact that, as it can be checked in a straightforward way, if $\ul x, \ul x' \in \SS_{\l_0,\e + \eta}$, $\l_0\in \WW_{\e,\t}$ and $\e,\eta$ are small enough there exists a constant $C_{\t}>0$ such that
\be
|x_{k}|\leq \frac{C_{\t}}{|\l_0|^{-1} + k}\;,\qquad \left|\frac{x_{k}}{x'_{h}}\right|\leq C_{\t}\;\quad \mbox{if $k\geq h$}\;;\label{A0}
\ee
the constant $C_{\t}$ grows as $\sim \t^{-1}$ for $\t\rightarrow 0$.
\vskip.2cm
{\bf Proof of item 1.} First, we have to prove that if $(\l_0,\a_0,\m_0)\in \WW_{\e,\t}\times \BB_{\eta}\times \BB_{\eta}$ and $\ul{x}\in \SS_{\l_0,\e+\eta}$ the map $\widetilde\TT_{\ul{\xi_0},\ul{x}}$ leaves invariant $\widetilde\SS_{2\eta}$, for $\t\in\big(0,\frac{\pi}{2}\big]$, and $\e$, $\eta$ small enough; in fact, setting $\ul a = (a_{1},...,a_{r})$, $\ul{h} = (h_{1},...,h_{r})$:
\bea
\Big|\left(\widetilde\TT_{\ul{\xi_0},\ul{x}}\,\ul{\ul{y}}\right)_{k,2'}\Big|&\leq&
|\a_0|+\sum_{j=1}^{k}\sum_{r\geq 2}\sum_{h_{i}\geq
j\atop i=1,\cdots,r}\sum_{\{a_{i}\}_{i=1}^{r}}\big|\beta_{\ul{a}}^{(2')}(j; \ul h)\big|\prod_{i = 1 \atop a_{i} = 4}^{r}|x_{h_{i}}|\prod_{i = 1 \atop a_{i}\neq4}^{r}|y_{h_{i},a_{i}}|\nn\\
&\leq&|\a_0|+\sum_{j=1}^{k}\sum_{r\geq 2}\sum_{h_{i}\geq
j\atop i=1,\cdots,r}\sum_{\{a_{i}\}_{i=1}^{r}}\big|\beta_{\ul{a}}^{(2')}(j;\ul{h})\big||x_j|^{2}C_{\t}^{\bar r}\varepsilon^{\bar r-2}(2\eta)^{\tilde{r}}\nn\\
&\leq&|\a_0| + C'_{\t}\e\label{eq:ac4}
\eea
for some $C'_{\t}>0$; similarly,
\bea
\Big|\left(\widetilde\TT_{\ul{\xi_0},\ul{x}}\,\ul{\ul{y}}\right)_{k,2}\Big|&\leq& |\m_0| + \sum_{j=1}^{k}\gamma^{2(j-k)}\sum_{r\geq 2}\sum_{h_{i}\geq
j\atop i=1,\cdots,r}\sum_{\{a_{i}\}_{i=1}^{r}}\big|\beta_{\ul{a}}^{(2)}(j;\ul h)\big|C_\t^{\bar r}\varepsilon^{\bar r}(2\eta)^{\tilde{r}}\nn\\
&\leq& |\m_0| + C'_{\t}(\varepsilon+\eta)^{2}\;.\label{eq:ac5}
\eea
for $C'_{\t}$ large enough. Hence, if $(\a_{0},\m_{0})\in \BB_{\eta}\times \BB_{\eta}$, then both (\ref{eq:ac4}), (\ref{eq:ac5}) can be made smaller than $2\eta$ taking $\varepsilon$ small enough.
\vskip.2cm
{\bf Proof of item 2.} Under the same assumption of item {\bf 1.}, we show now that $\widetilde\TT_{\ul{\xi_0},\ul x}$ is a contraction in $\widetilde\SS_{2\eta}$; in fact,
\bea
&&\Big|\left(\widetilde\TT_{\ul{\xi_0},\ul{x}}\,\ul{\ul{y}}\right)_{k,2'}-\left(\widetilde\TT_{\ul{\xi_0},\ul{x}}\,\ul{\ul{y}}'\right)_{k,2'}\Big|\leq\sum_{j=1}^{k}\sum_{r\geq
3\atop \{a_{i}\}_{i=1}^{r}}\sum_{h_{i}\geq j\atop i=1,...,r}\big|\beta_{\ul{a}}^{(2')}(j;\ul{h})\big| \prod_{i = 1 \atop a_{i} = 4}^{r}|x_{h_{i}}|\cdot\lb{ac5b}\\&&\hskip8cm \cdot (6\eta)^{\tilde r - 1}\tilde r\max_{k,i}\big| y_{k,i} - y'_{k,i} \big|\;,\nn
\eea
where we used that the second order of the beta function depends only on $x_{j}$; therefore, we can exploit two of the $x_{h_i}$'s to perform the sum, and it follows that, for $\e$, $\eta$ small enough:
\bea
\left|\Big(\widetilde\TT_{\ul{\xi_0},\ul{x}}\,\ul{\ul{y}}\right)_{k,2'}-\left(\widetilde\TT_{\ul{\xi_0},\ul{x}}\,\ul{\ul{y}}'\Big)_{k,2'}\right|&\leq&\epsilon \max_{k,i}|y'_{k,i}-y_{k,i}|\;,\qquad 0<\epsilon<1\;.
\eea
The same result can be proved for the difference of the $2$-components, using the $\g^{2(j-k)}$ factor to perform the sum over the $j$'s; this concludes the proof of the contractivity of $\widetilde\TT_{\ul{\xi_0},\ul x}$.
\vskip.2cm
{\bf Proof of item 3.} We prove here the last item of lemma \ref{lem:1}. Given $\ul{\ul{y}}\in \widetilde\SS_{2\eta},$ set
\be
y_{k,i,n}:= \Big({\widetilde\TT}^{n}_{\ul{\xi_0},\ul{x}}\,\ul{\ul{y}}\Big)_{k,i}\;,\qquad y'_{k,i,n}:= \Big({\widetilde\TT}^{n}_{\ul{\xi_0},\ul{x}'}\,\ul{\ul{y}}\Big)_{k,i}\;,\label{i3.1}
\ee
and assume inductively that for all $0\leq m\leq n$ the following bound is true:
\be
\big| y_{k,i,m} - y'_{k,i,m} \big| \leq C\big(\log(1+\e k) + 1\big)\max_k\big|x_k - x'_k\big|\;;\label{i3.2}
\ee
therefore, from (\ref{i3.2}) it follows that:
\bea
\left|y_{k,2',n+1}-y'_{k,2',n+1}\right|&\leq& \sum_{j=1}^{k}\sum_{r\geq
2 \atop \{a_{i}\}_{i=1}^{r}}\sum_{h_{i}\geq
j \atop i=1,\cdots r}\big|\beta^{(2')}_{\ul a}(j;\ul h)\big| \Big[C_{\t}^{\bar r-1} |x_j| \bar r (3\varepsilon)^{\bar r-2}(6\eta)^{\tilde{r}} + \lb{i3.3}\\&& +
\sum_{\ell = 1\atop a_{\ell}\neq 4}^{r}CC_{\t}^{\bar r}\big( \log(1+\varepsilon h_{\ell}) +
1\big) |x_j|^2 (3\varepsilon)^{\bar r-2} (6\eta)^{\tilde{r}-1}
\Big] \max_{k}|x_{k}-x'_{k}|\;,\nn
\eea
and
\bea
\big|y_{k,2,n+1}-y'_{k,2,n+1}\big|&\leq&\sum_{j=1}^{k}\gamma^{2(j-k)}\sum_{r\geq
2 \atop \{a_{i}\}_{i=1}^{r}}\sum_{h_{i}\geq
j\atop i=1,\cdots,r}\big|\beta^{(2)}_{\ul a}(j;\ul h) \big|\Big[C_\t^{\bar r-1}\bar r (3\varepsilon)^{\bar r-1}(6\eta)^{\tilde{r}} + \label{i3.4}\\&& \hskip1.5cm +
\sum_{\ell = 1 \atop a_{\ell}\neq 4}^{r}CC_{\t}^{\bar r}\big(\log(1+\varepsilon h_{\ell}) +
1\big) (3\varepsilon)^{\bar r}(6\eta)^{\tilde{r}-1}
\Big] \max_{k}|x_{k}-x'_{k}|.\nn
\eea
Using the short memory property of the GN trees, see discussion after (\ref{1.3.7}), it follows that:
\bea
\sum_{\substack{h_{1},\ldots , h_{r} \\ h_{i}\geq j}} \big| \b^{(i)}_{\ul a}(j;\ul h) \big|\log(1 + \e h_{\ell}) \leq (\const.)^r \log(1 + \e j)\;;
\eea
plugging this bound into (\ref{i3.3}), (\ref{i3.4}) we can reproduce our inductive assumption (\ref{i3.2}) for $m=n+1$, choosing for $\e$, $\eta$ small enough. This concludes the proof of
(\ref{eq:reg1}). The ``H\"older continuity bound'' (\ref{reg1b}) can be proved again by induction, replacing (\ref{i3.2}) with
\be
\big| y_{k,i,m} - y'_{k,i,m} \big| \leq C_\r \big(\max_k |x_k - x'_k|\big)^{\r}\;,\qquad 0<\r<1\;,\label{i3.6}
\ee
and using in (\ref{i3.3}) the bound, if $h_{i}\geq j$ for all $i=1,\ldots,r$ and $0<\r<1$:
\be
\Big| \prod_{i=1 \atop a_i = 4}^{r}x_{h_i} - \prod_{i=1\atop a_i = 4}^{r}x'_{h_i} \Big| \leq 2\bar r \big(\max_{k}|x_k - x'_k|\big)^{\r}|x_j|^{2-\r}C^{\bar r}_{\t}(3\e)^{\bar r -2}\;.\label{i3.7}
\ee

\section{Proof of lemma \ref{lem:2}}\label{app:B}
\renewcommand{\theequation}{\ref{app:B}.\arabic{equation}}

In this appendix we present a proof of lemma \ref{lem:2}.

\vskip.2cm
{\bf Proof of item 1.} First, we have to prove that $\TT_{\l_{0},\ul{\ul y}(\cdot )}$ leaves invariant $\SS_{\l_0,\e + \eta}$ for $(\l_0,\a_0,\m_0)\in \WW_{\e,\t}\times \BB_{\eta}\times \BB_{\eta}$ for $\e,\eta$ small enough. We have that
\be
\left(\TT_{\l_{0},\ul{\ul y}(\ul x)} \ul{x}\right)_{k}=\frac{1}{\l_0^{-1}+\sum_{j=1}^{k}\b_{j}
-\sum_{j = 1}^{k}R_{j}\big(\underline{x},\ul{\ul{y}}(\ul x)\big)}\;,\label{B0.0}
\ee
where $\ul{\ul{y}}(\ul x)\in\widetilde\SS_{2\eta}$ and
\be
R_{j}\big(\ul{x},\ul{\ul{y}}(\ul x)\big) =
\frac{x_{j}\b_{j}^{2}+x_{j}^{-1}\b_{j}\bar\b_{4,j}\big(\ul{x},\ul{\ul{y}}(\ul x)\big)-x_{j}^{-2}\bar\b_{4,j}\big(\ul{x},\ul{\ul{y}}(\ul x)\big)}{1+\b_{j}x_{j}+x_{j}^{-1}\bar\b_{4,j}\big(\ul{x},\ul{\ul{y}}(\ul x)\big)}\;,\lb{B0}
\ee
with
\be
\bar\b_{4,j}\big(\ul{x},\ul{\ul{y}}(\ul x)) = \sum_{r\geq 3}\sum_{h_{i}\geq
j\atop i=1,\cdots,r}\sum_{\{a_{i}\}_{i=1}^{r}}\beta_{\ul a}^{(4)}(j;\ul h)\prod_{i = 1\atop a_{i}=4}^{r}x_{h_{i}}\prod_{i = 1\atop a_{i}\neq
4}^{r}y_{h_{i},a_{i}}(\ul x)\;,
\ee
where $\beta_{a_1,...,a_r}^{(4)}(j;h_1,...,h_r)=0$ unless there are at least two
$a_{i}$ equal to $4$. The final statement follows from the fact that
for $\varepsilon$, $\eta$ small enough
\be
\big|R_{j}(\underline{x},\ul{\ul{y}}(\ul x))\big|\leq (\const.)(\varepsilon+\eta)\;.
\ee
\vskip.2cm
{\bf Proof of item 2.} To conclude, we have to show that under the same assumptions of the previous item, $\TT_{\l_{0},\ul{\ul y}(\ul x)}$ is a contraction in $\SS_{\l_0,\e + \eta}$. Setting $\ul{\ul{y}}(\underline{x})=:\ul{\ul{y}}$,
$\ul{\ul{y}}(\underline{x}')=: \ul{\ul{y}}'$, from (\ref{B0.0}) we have
that
\bea
&&\left(\TT_{\l_{0},\ul{\ul y}(\ul x)}\ul{x}\right)_k-\left(\TT_{\l_{0},\ul{\ul y}(\ul x')}\ul{x}'\right)_k = \label{eq:B2.1}\\&&\hskip1cm=\frac{\sum_{j=1}^{k}\left(R_{j}(\ul{x},\ul{\ul{y}})-R_{j}(\ul{x}',\ul{\ul{y}'})\right)}{\left(\l_0^{-1}+\sum_{j=1}^{k}\b_{j}
-\sum_{j =
1}^{k}R_{j}(\underline{x},\ul{\ul{y}})\right)\left(\l_0^{-1}+\sum_{j=1}^{k}\b_{j}
-\sum_{j = 1}^{k}R_{j}(\underline{x}',\ul{\ul{y}}')\right)}\;,\nn
\eea
where $R_j$ is given by (\ref{B0}); therefore, to bound the difference of $R_{j}$'s calculated at
different $\underline{x}$ we have to estimate (the other terms can be worked
out in a similar way)
\bea
&&\Big|x_{j}^{-2}\bar\b_{4,j}\big(\ul{x},\ul{\ul{y}}\big)-{x'_{j}}^{-2}\bar\b_{4,j}\big(\ul{x}',\ul{\ul{y}}'\big)\Big|\leq
\sum_{r\geq 3}\sum_{\{h_{i}\}\geq
j\atop \{a_{i}\}}\big |\beta_{\ul a}^{(4)}(j;\ul h)|\cdot \nn\\&&\hskip3cm \cdot \Big|x_{j}^{-2}\prod_{i = 1\atop a_{i}=4}^{r}x_{h_{i}}\prod_{i = 1\atop a_{i}\neq
4}^{r}y_{h_{i},a_{i}}-{x'_{j}}^{-2}\prod_{i = 1\atop a_{i}=4}^{r}x'_{h_{i}}\prod_{i = 1\atop a_{i}\neq
4}^{r}y'_{h_{i},a_{i}}\Big|\;;\label{eq:lam1}
\eea
we have that
\bea
&&\Big|x_{j}^{-2}\prod_{i = 1\atop a_{i}=4}^{r}x_{h_{i}}\prod_{i = 1\atop a_{i}\neq
4}^{r}y_{h_{i},a_{i}}-{x'_{j}}^{-2}\prod_{i = 1\atop a_{i}=4}^{r}x'_{h_{i}}\prod_{i = 1\atop a_{i}\neq
4}^{r}y'_{h_{i},a_{i}}\Big|\leq \nn\\&&\hskip2cm \leq
\Big|x_{j}^{-2}\Big(\prod_{i = 1\atop a_{i}=4}^{r}x_{h_{i}}\prod_{i = 1\atop a_{i}\neq
4}^{r}y_{h_{i},a_{i}}-\prod_{i = 1\atop a_{i}=4}^{r}x'_{h_{i}}\prod_{i = 1\atop a_{i}\neq
4}^{r}y'_{h_{i},a_{i}}\Big)\Big| + \lb{eq:lam1c}\\&&\hskip3cm +\max_{k}|x_{k}-x'_{k}|\Big|\frac{(x'_{j}+x_{j})}{x_{j}^{2}{x'_{j}}^{2}}\prod_{i = 1\atop a_{i}=4}^{r}x'_{h_{i}}\prod_{i = 1 \atop a_{i}\neq
4}^{r}y'_{h_{i},a_{i}}\Big|\label{eq:lam1b}
\eea
and:
\bea
&&(\ref{eq:lam1c})\leq
|x_{j}|^{-1}C_{\t}^{\bar r-1} r(3\e)^{\bar r - 2}(6\eta)^{\tilde r}\max_{k}|x_{k}-x'_{k}| + \nn\\&&\hskip6cm +
C_{\t}^{\bar r}(3\e)^{\bar r-2}(6\eta)^{\tilde r - 1}\sum_{\ell = 1\atop a_\ell \neq
4}^{r}\left|y_{h_{\ell},a_{\ell}}-y'_{h_{\ell},a_{\ell}}\right|\qquad\label{C.0.1}\\
&&(\ref{eq:lam1b})\leq
\max_{k}|x_{k}-x'_{k}| |x_{j}|^{-1}2C_{\t}^{\bar r}(3\varepsilon)^{\bar r-2}(6\eta)^{\tilde{r}}\;.\qquad\label{C.0.2}
\eea
Using (\ref{eq:reg1}) and the short memory property it follows that:
\be
\sum_{\substack{h_1,\ldots ,h_{r}\\ h_{i}\geq j}}\big| \b^{(4)}_{\ul a}(j;\ul h) \big|\big| y_{h_{\ell},a_{\ell}} - y'_{h_{\ell},a_{\ell}} \big|\leq (\const.)^r \big[\log(1 + \e j) + 1\big]\max_{k}|x_k - x'_k|\;;\label{C.0.3}
\ee
therefore, since the other terms arising in the difference (\ref{eq:B2.1}) can be treated exactly in the same way, from (\ref{C.0.1})--(\ref{C.0.3}) we find that:
\be
\sum_{j=1}^{k}\Big|R_{j}(\underline{x}',\ul{\ul{y}}')-R_{j}(\underline{x},\ul{\ul{y}})\Big| \leq (\const.)\Big[\sum_{j=1}^{k}|x_j|^{-1}(\e + \eta) + k\big(\log(1 + \e k) +1\big)\big]\max_{k}|x_k - x'_k|\;,
\ee
which gives statement (\ref{2.6}) for $\varepsilon$, $\eta$ small enough. In fact, the denominator of (\ref{eq:B2.1}) is bounded from below as
\bea
&&\Big|\l_0^{-1}+\sum_{j=1}^{k}\Big(\b_{j} - R_{j}(\underline{x},\ul{\ul{y}})\Big)\Big|\Big|\l_0^{-1}+\sum_{j=1}^{k}\Big(\b_{j}
- R_{j}(\underline{x}',\ul{\ul{y}}')\Big)\Big| \geq (\const.)|x_k|^{-2}\;;
\eea
using the second of (\ref{A0}) our claim (\ref{2.6}) follows.

\section{The running coupling constants on scale $0$.}\lb{appscale0}
\renewcommand{\theequation}{\ref{appscale0}.\arabic{equation}}

In this appendix we discuss how to express the running coupling constants on scale zero as functions of the renormalized ones. First, a straightforward computation shows that the second equation in (\ref{1.8b}) can be rewritten as:
\be
\m_0 = \frac{\m}{1+\m} - \frac{1-\m_0}{1+\m}\b_{2,0}(\l_0,\ul{\l},\ul{\xi_0},\ul{\ul \xi})\;;\label{2.8.0}
\ee
this is a consequence of the fact that in (\ref{1.4}) $\b^{(2)}_{a_1,\ldots, a_{r}}(0;0\ldots , 0) = 1$ if $a_{i} = 2$ for all $i\in[1,r]$. Since the running coupling constants on scale $>0$ are parametrized by the ones on scale $0$, we can rewrite (\ref{2.8.0}) as
\be
\m_0 =: \frac{\m}{1 + \m} + g_{2}(\l_0,\ul{\xi_0},\m) =: \m + \tilde f_{2}(\m) + g_2(\l_0,\ul{\xi_0},\m)\;,\label{2.8.1}
\ee
and plugging (\ref{2.8.1}) in the first equation of (\ref{1.8b}) we get
\be
\a_{0} =: \a + \tilde f_{2'}(\m) + g_{2'}(\l_0,\ul{\xi_0},\m)\;, \label{2.8.2}
\ee 
where: $\tilde f_{i}(\m)$ are analytic functions of $\m\in\BB_{\bar\eta}$, and $g_{i}(\l_0,\ul{\xi_0},\m)$ are analytic for $(\l_{0},\ul{\xi_0},\m)\in \WW_{2\bar \e,\frac{\t}{2}}\times \BB_{2\bar \eta}\times \BB_{2\bar\eta}\times \BB_{\bar\eta}$. Formulas (\ref{2.8.1}), (\ref{2.8.2}) can be regarded as a fixed point equation:
\be
\xi_{i,0} = \Big(\widetilde\MM_{\ul\xi,\l_0}\,\ul{\xi_0}\Big)_{i}\;;\lb{2.8d}
\ee
all we have to do is to check that: (i) for $|\l_0|,|\ul\xi|$ small enough $\widetilde\MM_{\ul\xi,\l_0}$ leaves invariant the set $\BB_{\frac{3}{2}\bar \eta}\times \BB_{\frac{3}{2}\bar \eta}$, and (ii) $\widetilde\MM_{\ul\xi,\l_0}$ is a contraction therein. The property (i) is a straightforward consequence of the fact that 
\be
\big|\tilde f_{i}(\m)\big|\leq C|\m|^2\,,\qquad \big| g_{i}\big(\l_{0},\ul{\xi_0},\m\big) \big|\leq C|\l_{0}|\Big(|\l_0| + |\ul\xi|\Big)\;,\lb{2.8e}
\ee
where in the second inequality we used that $|\l_k|\leq c|\l_0|$, $|\xi_{i,k}|\leq c\big( |\l_0| + |\ul{\xi_0}| \big)$ and that, from (\ref{1.8b}), $|\xi_{i,0}|\leq c\big( |\l_0| + |\ul\xi| \big)$; if we choose $(\l_0,\ul{\xi})\in \WW_{2\bar\e,\frac{\t}{2}}\times\BB_{\bar\eta}\times\BB_{\bar\eta}$ with $\bar\e$, $\bar\eta$ small enough then the set $\BB_{\frac{3}{2}\bar\eta}\times\BB_{\frac{3}{2}\bar\eta}\subset \BB_{2\bar \eta}\times \BB_{2\bar \eta}$ is left invariant by (\ref{2.8d}).

To prove property (ii) we use a Cauchy estimate. In fact, the Cauchy bound tells us that if $\ul{y},\ul{y}'\in \BB_{\frac{3}{2}\bar\eta}\times\BB_{\frac{3}{2}\bar\eta}$ then, since $g_{i}(\l_0,\ul y,\mu)$ is analytic for $\ul y\in \BB_{2\bar\eta}\times\BB_{2\bar\eta}$ and bounded as (\ref{2.8e}), for $(\l_{0},\m)\in \WW_{2\bar\e,\frac{\t}{2}}\times \BB_{\bar\eta}$ with $\bar\e, \bar\eta$ small enough:
\be
\big| g_{i}(\l_0,\ul{y}, \m) - g_{i}(\l_0,\ul{y}', \m) \big| \leq 2C\frac{\bar\e}{\bar\eta}\big(\bar\e + \bar\eta\big)\max_{i}|y_{i} - y'_{i}| \leq \epsilon \max_{i}|y_{i} - y'_{i}|\;\lb{2.8f}
\ee
with $0<\epsilon<1$. Therefore, we can construct explicitly the solution $\xi_{i,0}(\l_0,\ul{\xi})$, and the above properties allow us to conclude that it is analytic for $(\l_{0},\ul\xi)\in \WW_{2\bar\e,\frac{\t}{2}}\times \BB_{\bar\eta} \times \BB_{\bar\eta}$.

After this, we are left with the equation (\ref{1.7a}) for $\l_{0}$; since all the couplings on scale $\geq 1$ are functions of $\l_{0}$, $\ul{\xi_0}$ and, as we know for our previous analysis, $\xi_{i,0} = \xi_{i,0}(\l_{0},\ul{\xi})$, we can rewrite (\ref{1.7a}) as:
\bea
&&\l_{0} =:  \l - \l_0 \tilde f_4(\m) - \b_{0}\l_0^{2} + h(\l_{0},\ul{\xi})\lb{2.8ga}\\ &&\tilde f_4(\m) = O(\m)\;,\qquad \big|h(\l_{0},\ul{\xi})\big| \leq C|\l_{0}|^{2}\Big(|\l_0| + |\ul\xi|\Big)\;,\nn
\eea
where we used that $\m_0$ satisfies (\ref{2.8d}) with $i=2$, and $h(\l_0,\ul{\xi})$ is analytic for $(\l_0,\ul{\xi})\in \WW_{2\bar\e,\frac{\t}{2}}\times\BB_{\bar\eta}\times \BB_{\bar\eta}$. Therefore, we can rewrite (\ref{2.8ga}) as:
\bea
&&\l_0 = \MM_{\tilde\l,\ul{\xi}}\,\l_0\;,\nn\\&&\tilde\l := \frac{\l}{1+\tilde f_4(\m)}\;,\quad \MM_{\tilde\l,\ul\xi}\, x := \tilde\l + \frac{1}{1+ \tilde f_4(\m)}\Big( -\b_0x^{2} + h(x,\ul{\xi}) \Big)\;.\label{2.8i}
\eea
All we have to do is to check that: (i) if $(\l,\ul{\xi})\in\WW_{\bar\e,\t}\times\BB_{\bar\eta}\times\BB_{\bar\eta}$ then $\MM_{\tilde\l,\ul{\xi}}$ leaves invariant the set $\WW_{\frac{3}{2}\bar\e,\frac{2}{3}\t}\subset \WW_{2\bar\e,\frac{\t}{2}}$, and (ii) $\MM_{\tilde\l,\ul{\xi}}$ is a contraction therein. Let us prove property (i); for $\bar\e$, $\bar\eta$ small enough, it is easy to see that if $\l\in\WW_{\bar\e,\t}$ then $\tilde{\l}\in \WW_{\frac{4}{3}\e,\frac{3}{4}\t}$ and $x\in \WW_{\frac{3}{2}\bar\e,\frac{2}{3}\t} \Rightarrow \MM_{\tilde \l,\ul{\xi}}\,x\in \WW_{\frac{3}{2}\bar\e,\frac{2}{3}\t}$.

We now turn to property (ii). From the analyticity of $h(x,\ul{\xi})$ in $x\in \WW_{2\bar\e,\frac{\t}{2}}$, using that the distance from a point $x\in \WW_{\frac{3}{2}\bar\e,\frac{2}{3}\t}$ to the boundary of $\WW_{2\bar\e,\frac{\t}{2}}$ is bounded from below by $\frac{|x|}{3}\sin\frac{\t}{6}$, if $x,x'\in \WW_{\frac{3}{2}\bar\e,\frac{2}{3}\t}$ a Cauchy estimate tells us that:
\be
\Big| \MM_{\tilde\l,\ul{\xi}}\,x  - \MM_{\tilde\l,\ul{\xi}}\,x'\big| \leq 8\Big[\b_0\bar\e + \frac{3C\bar\e(\bar\e + \bar\eta)}{\sin(\t/6)} \Big]|x - x'|\leq \epsilon |x - x'|\label{2.8ia}
\ee
with $\epsilon <1$; the first inequality follows from the bound on $h$ in (\ref{2.8ga}), while the second holds taking $\bar\e$ small enough (remember that $\t\in\big(0,\frac{\pi}{2}\big]$).

In conclusion, we can explicitly construct the solution of (\ref{2.8i}), and by a simple inductive argument it follows that it is analytic for $(\l,\a,\m)\in \WW_{\bar\e,\t}\times\BB_{\bar\eta}\times\BB_{\bar\eta}$.

\section{Dependence of the running coupling constants on the ultraviolet cutoff}\lb{appuv}
\renewcommand{\theequation}{\ref{appuv}.\arabic{equation}}

In this appendix we show that the running coupling constants are weakly dependent on the location of the ultraviolet cutoff; in particular, denoting with a superscript $N$ the quantities corresponding to a theory with cutoff on scale $N$, if $(\l,\a,\m)\in \WW_{\e,\t}\times \BB_{\eta}\times \BB_{\eta}$ with $\e$, $\eta$ small enough we show that there exist two positive constants $C,\r$ such that for any $k\leq N$ and $N<N'$ the following bounds hold:
\bea
&&\big| \a_{k}^{N} - \a_{k}^{N'} \big| \leq \frac{C}{\e^{-1} + N} + C\eta\g^{-\r N}\;,\quad \big| \m_{k}^{N} - \m_{k}^{N'} \big|\leq \frac{C}{\e^{-1} + N} + C\eta\g^{-\r N},\nn\\&&\big| \l_{k}^{N} - \l_{k}^{N'} \big| \leq \frac{C}{\e^{-1} + N}\;.\lb{D1}
\eea
In the proof we shall use in a crucial way the short memory property of the GN trees, see discussion after (\ref{1.3.7}). Consider first the difference of $\a_{k}^{N}$, $\a_{k}^{N'}$. Denoting by a prime the running coupling constants corresponding to a theory with cutoff $N'$ and neglecting the $N$ label in the others we have that
\be
\big| \a_k - \a'_k \big| \leq \sum_{j=1}^{k}\big| \b_{2',j}^{N}(\ul\l,\ul\a,\ul\m) - \b_{2',j}^{N'}(\ul\l',\ul\a',\ul\m') \big| + |\a_0 - \a_0'|\;,\lb{D2}
\ee
where $\b_{2',j}^{N}$ is the beta function the theory with an ultraviolet cutoff on scale $N$. Let $\|a\| := \max_{k\in [0,N]}|a_k|$; using property (\ref{1.4b}) and the bounds in (\ref{A0}) it follows that, for some $C_1 > 0$, $\r>0$ (neglecting for simplicity the arguments of the beta function):
\bea
\big| \b_{2',j}^{N} -  \b_{2',j}^{N'}\big| &\leq& \frac{C_1}{(\e^{-1} + j)^{2}}\big(\|\a - \a'\| + \| \m - \m' \|\big) + \frac{C_1}{\e^{-1} + j}\sum_{h\geq j}|\l_h - \l_h'|\g^{\r(j - h)} +\nn\\&&+ C_1(\e + \eta)\frac{\g^{\r(j - N)}}{\e^{-1} + j}\;,\label{D2b}\\
|\a_0 - \a'_0| &\leq& C_{1}(\e + \eta)\Big( \|\a - \a'\| + \|\m - \m'\| + \|\l - \l'\| \Big) + C_{1}(\e+\eta)^2\g^{-\r N}\;,\label{D2c}
\eea
where the last terms in (\ref{D2b}), (\ref{D2c}) take into account the contribution of GN trees with at least one endpoint on scale $> N$, and all the others bound the differences of trees with all endpoints on scale $<N$. Therefore, plugging (\ref{D2b}), (\ref{D2c}) in (\ref{D2}) we have that, for some $\tilde C_{1}>0$:
\bea
\| \a - \a' \| &\leq& \tilde C_1(\e+\eta) \big(\|\m - \m'\| + \|\l - \l'\|\big) + \sum_{j=1}^{N}\frac{\tilde C_1}{\e^{-1}+j}\sum_{h\geq j}^N|\l_h - \l'_h| \g^{\r(j-h)}+\nn \\&& + \frac{\tilde C_1(\e + \eta)}{\e^{-1} + N} + \tilde C_1(\e + \eta)^2\g^{-\r N}\;.\lb{D4}
\eea
By what has been discussed in sections \ref{sec3}, \ref{sec4} and in appendices \ref{app:A}, \ref{app:B}, it follows that $|\l_k - \l_k'|$ for $k\geq 1$ can be estimated in the following way, for some $C_{2}>0$:
\be
|\l_{k} - \l_{k}'| \leq \frac{C_2\big( \|\l - \l'\| + \|\a - \a'\| + \|\m - \m'\| \big)}{\e^{-1} + k} + \frac{C_2(\e + \eta)\g^{\r(k - N)}}{(\e^{-1} + k)^2}\;,\lb{D6}
\ee
where the first term takes into account the difference of running coupling constants on scale $\leq N$, while the last term takes into account trees with root scale $\leq k$ having at least one endpoint on scale $> N$. Plugging (\ref{D6}) in (\ref{D4}) it is straightforward to see that, for some $C_3>0$,
\be
\| \a - \a' \|\leq C_3 (\e + \eta)\big(\| \l - \l' \| + \| \m - \m' \|\big) + \frac{C_3 (\e + \eta)}{\e^{-1} + N} + C_3(\e + \eta)^2\g^{-\r N}\;,\lb{D7}
\ee
which if inserted in (\ref{D6}) implies, for some positive $C_4, C_5$:
\bea
\|\l - \l'\| &\leq& \e C_4 \|\m - \m'\| + C_4\e(\e + \eta)^2 \g^{-\r N} + \frac{C_4\e(\e + \eta)}{\e^{-1} + N}\;,\nn\\
\|\a - \a'\| &\leq& C_5(\e + \eta) \| \m - \m' \| + \frac{C_5(\e + \eta)}{\e^{-1} + N} + C_5 (\e + \eta)^2 \g^{-\r N}\;.\label{D8}
\eea
The difference $\m_{k} - \m'_{k}$ can be bounded in a way analogous to $\a_k-\a'_k$, and using (\ref{D8}) it follows that
\be
\|\m - \m'\|\leq \frac{C_6}{\e^{-1} + N} + C_6\eta\g^{-\r N}\;,\qquad C_{6}>0\;,\lb{D9}
\ee
which together with (\ref{D8}) proves (\ref{D1}).

\section{An improvement of the $n!$ bounds in the planar theory}\lb{appD}
\renewcommand{\theequation}{\ref{appD}.\arabic{equation}}

In this appendix we discuss an improvement, valid in the planar case, 
of the $n!$ bounds proved in \cite{G} in section XIX, see formulas (19.5), (20.2). Here we shall follow the notations of that work:
we remind that the ``form factor'' $r^{(a)}(\s;k)$ of \cite{G} corresponds to the contribution of the tree $\s$ 
with thin endpoints to the formal expansion of
$v_k^{(a)} \g^{(2\d_{a,2}+4\d_{a,0})k}$ in $\l,\a,\m$, which is obtained by iteration of the 
equation graphically represented in figure \ref{fig03}. We claim that equation \cite{G}--(20.2) is still valid if $f$ is replaced by an $\bar{f}$
denoting just the number of nontrivial frames (see remark after figure \ref{fig03} for the definition of trivial frame) labelled by $a=2',4$. 

To prove the claim, observe that one can repeat the proof of section XIX in \cite{G} with the new inductive assumption
\be
|r^{(a)}(\s;k)| \leq \bar{\e}^n \tilde{D}^{n-1} \bar{f}! \sum_{j=0}^{\bar{f}}\frac{(bk)^j}{j!}\g^{(2\d_{a,2}+4\d_{a,0})k}
\ee
instead of \cite{G}--(19.5), the only difference being that the number of topological Feynman
graphs with $m$ vertices is bounded proportionally to $N_0^m$ where $N_0$ is a suitable constant, because 
of the restriction to the planar theory.
Then if $\tilde f$ is the number of nontrivial $(2',4)$--frames of $\s$ {\em excluding} the external one, equation \cite{G}--(19.13) is replaced by, depending on whether the frame enclosing $\s$ is trivial or not:
\bea
|r^{(a)}(\s;k)| &\leq& D_7 N_0^m D_4^m \bar{\e}^n \tilde{D}^{n-m} D_6^m
\tilde f! \sum_{h=0}^k\sum_{r=0}^{\tilde f}
\g^{(2\d_{a,2}+4\d_{a,0})h}\frac{(bh)^r}{r!}\;,\quad\mbox{(non trivial frame),}\nn\\
|r^{(a)}(\s;k)| &\leq& D_7 N_0^m D_4^m \bar{\e}^n \tilde{D}^{n-m} D_6^m
\tilde f!\;,\hskip3.5cm \mbox{(trivial frame);}\label{f0}
\eea
with respect to \cite{G}, we have kept the factor $\g^{(2\d_{a,2}+4\d_{a,0})h}$ inside the sum, instead of estimating it replacing $h$ with $k$. If the frame enclosing $\s$ is trivial the claim follows from the second of (\ref{f0}), taking $\tilde D$ large enough (as in \cite{G}, here $m\geq 2$). If the frame is nontrivial and $a = 2',4$, proceed as in \cite{G}--(19.15), while if $a = 2$ substitute that bound with
\be
\sum_{h=0}^k\sum_{r=0}^{\bar{f}}
\g^{2h}\frac{(bh)^r}{r!} \leq \frac{\g^{2k}}{1-\g^{-2}} \sum_{r=0}^{\bar{f}}\frac{(bk)^r}{r!}\;,
\ee
and do the same for $a=0$ ($\g^{2k}$ will be replaced by $\g^{4k}$). From this the claim follows choosing $\tilde{D}$ sufficiently large, as explained in \cite{G}.


\begin{thebibliography}{99999999999999999999999999999}

\bibitem{L}
L. D.~{Landau}.
\newblock{\em Collected papers of L. D. Landau}.
\newblock Gordon and Breach, 1965.

\bibitem{tH}
G.~{'t Hooft}.
%\newblock {Borel summability of a four -- dimensional field theory}.
\newblock {\em {Phys. Lett. B}} {\bf 119}, 369--371 (1982).

\bibitem{tH2}
G.~{'t Hooft}.
%\newblock {Rigorous construction of planar diagram field theories in four
  %dimensional euclidean space.}
\newblock {\em {Comm. Math. Phys.}} {\bf 88}, 1--25 (1983).

\bibitem{R}
V.~{Rivasseau}.
%\newblock {Construction and Borel summability of planar 4 -- dimensional
  %Euclidean field theory}.
\newblock {\em {Comm. Math. Phys.}} {\bf 95}, 445--486 (1984).

\bibitem{R2}
V.~{Rivasseau}.
%\newblock {Rigorous construction and Borel summability for a planar four -- dimensional field theory}.
\newblock {\em {Phys. Lett. B}} {\bf 137}, 98--102 (1983).

\bibitem{G}
G.~{Gallavotti}.
%\newblock Renormalization theory and ultraviolet stability for scalar fields
  %via renormalization group methods.
\newblock {\em Rev. Mod. Phys.} {\bf 57}, 471--562 (1985).

\bibitem{GR}
G.~{Gallavotti} and V.~{Rivasseau}.
%\newblock {$\varphi^4$ -- Field Theory in dimension four: a modern introduction
  %to its open problems.}
\newblock {\em {Ann. Inst. H. Poincar\'e}}  {\bf 40}, 185--220 (1984).

\bibitem{FMRS}
F.~{Feldman}, J.~{Magnen}, V.~{Rivasseau} and R.~{S\'en\'eor}.
%\newblock{Construction and Borel summability of infrared $\Phi^4_4$ by a phase space expansion.}
\newblock{\em Comm. Math. Phys} {\bf 109}, 437--480 (1987).

\bibitem{FMRS2}
F.~{Feldman}, J.~{Magnen}, V.~{Rivasseau} and R.~{S\'en\'eor}.
%\newblock{A renormalizable field theory: the massive Gross--Neveu model in two dimensions.}
\newblock{\em Comm. Math. Phys.} {\bf 103}, 67--103 (1986).

\bibitem{K}
J.~{Koplik}, A.~{Neveu}, and {Nussinov} S.
%\newblock {Some aspects of the planar perturbation series}.
\newblock {\em {Nucl. Phys. B}} {\bf 123}, 109 (1977).

\bibitem{BIPZ}
E.~{Br\' ezin}, C.~{Itzykson}, G.~{Parisi}, and J.B. {Zuber}.
%\newblock {Planar diagrams}.
\newblock {\em {Comm. Math. Phys.}} {\bf 59}, 35--51 (1978).

\bibitem{GN1}
G.~{Gallavotti} and F.~{Nicol\`o}.
%\newblock {Renormalization theory for four dimensional scalar fields, I}.
\newblock {\em {Comm. Math. Phys.}} {\bf 100}, 545--590 (1985).

\bibitem{GN2}
G.~{Gallavotti} and F.~{Nicol\`o}.
%\newblock {Renormalization theory for four dimensional scalar fields, II}.
\newblock {\em {Comm. Math. Phys.}} {\bf 101}, 247--282 (1985).

\bibitem{S}
A.~{Sokal}.
%\newblock {An improvement of Watson's theorem on Borel summability}.
\newblock {\em J. Math. Phys.} {\bf 21}, 261 (1980).

\bibitem{M}
V.~{Mastropietro}.
%\newblock {Rigorous proof of Luttinger liquid behaviour in the $1d$ Hubbard
  %model}.
\newblock {\em {J. Stat. Phys.}} {\bf 121}, 373--432 (2005).

\bibitem{Hdivser}
G.~H. {Hardy}.
\newblock {\em Divergent Series}.
\newblock Oxford U. P., 1949.

\bibitem{R3}
V.~{Rivasseau}.
%\newblock {Constructive field theory in zero dimension}.
\newblock {\em Adv. Math. Phys.} {\bf 2009}, (2009).

\bibitem{Watson}
G.~N. {Watson}.
%\newblock{A theory of asymptotic series}.
\newblock {\em Philos. Trans. R. Soc. Lond. Ser. A} {\bf 211}, 279 (1912)

\bibitem{BG}
G.~{Benfatto} and G.~{Gallavotti}.
%\newblock {Perturbation theory of the Fermi surface in a quantum liquid. A general quasi--particle formalism and one dimensional systems}.
\newblock {\em{Comm. Math. Phys.}} {\bf 258}, 609--655 (2005).

\bibitem{GM}
G.~{Gentile} and V.~{Mastropietro}.
%\newblock {Renormalization group for one--dimensional fermions. A review on mathematical results}.
\newblock {\em{Phys. Rep.}} {\bf 352}, 273--437 (2001).

\bibitem{BM} 
G.~{Benfatto} and V.~{Mastropietro}.
%\newblock{Ward identities and chiral anomaly in the Luttinger liquid}.
\newblock {\em {Comm. Math. Phys.}} {\bf 258}, 609--655 (2005).

\bibitem{DR} 
C.~{De Calan} and V.~{Rivasseau}.
%\newblock{Local existence of the
%Borel transform in euclidean $\varphi^{4}_4$}.
\newblock {\em {Comm. Math. Phys.}} {\bf 82}, 69--100 (1982).

\bibitem{H}
K.~{Hepp}.
%\newblock{Proof of the Bogoliubov-Parasiuk theorem
%on renormalization}.
\newblock {\em {Comm. Math. Phys.}} {\bf 2}, 301--326 (1966).

\bibitem{Z}
W.~{Zimmermann}.
%\newblock{Convergence of Bogoliubov's method
%of renormalization in momentum space}.
\newblock{\em {Comm. Math. Phys.}} {\bf 15}, 208--234 (1969).

\bibitem{P}
J.~{Polchinski}.
%\newblock{Renormalization and effective Lagrangians}.
\newblock{\em {Nucl. Phys. B}} {\bf 231}, 269--295 (1984).

\end{thebibliography}
\end{document}